\documentclass[twocolumn,trackchanges]{aastex63}

\newcommand\kms{km\,s$^{-1}$}
\newcommand\msol{\mathrm{M}_\odot}
\newcommand\spy{\;\msol~\mathrm{ yr}^{-1}}

\definecolor{Leen}{rgb}{0.08, 0.69, 0.1} 
\definecolor{Carl}{rgb}{0.66, 0.37, 0.92} 
\definecolor{Anita}{rgb}{0,0,1}

\received{6 June 2020}
\revised{\today}
\accepted{}
\shorttitle{Rotational spectra of vibrationally excited AlO and TiO in oxygen rich stars}
\shortauthors{Danilovich, Gottlieb, Decin, Richards, Lee, Kaminski, Patel, Young and Menten}
\graphicspath{{./}{figures/}}

\begin{document}

\title{Rotational spectra of vibrationally excited AlO and TiO in oxygen rich stars}

\correspondingauthor{Ta\"issa Danilovich}\email{taissa.danilovich@kuleuven.be}

\author[0000-0002-1283-6038]{T. Danilovich}
\affiliation{Institute of Astronomy, KU Leuven, Celestijnenlaan 200D, 3001 Leuven, Belgium}

\author[0000-0003-2845-5317]{C. A. Gottlieb}
\affiliation{Harvard-Smithsonian Center for Astrophysics, 60 Garden Street, Cambridge, MA 02138, USA}

\author[0000-0002-5342-8612]{L. Decin}
\affiliation{Institute of Astronomy, KU Leuven, Celestijnenlaan 200D, 3001 Leuven, Belgium}

\author[0000-0002-3880-2450]{A. M. S. Richards}
\affiliation{JBCA, Department Physics and Astronomy, University of Manchester, Manchester M13 9PL, UK}

\author[0000-0002-1903-9242]{K. L. K. Lee}
\affiliation{Harvard-Smithsonian Center for Astrophysics, 60 Garden Street, Cambridge, MA 02138, USA}

\author[0000-0001-8541-8024]{T. Kami\'nski}
\affiliation{Nicolaus Copernicus Astronomical Center, Polish Academy of Sciences, Rabia\'nska 8, 87-100 Toru\'n, Poland}

\author[0000-0002-6021-9421]{N. A. Patel} 
\affiliation{Harvard-Smithsonian Center for Astrophysics, 60 Garden Street, Cambridge, MA 02138, USA}

\author[0000-0002-3666-4920]{K. H. Young} 
\affiliation{Harvard-Smithsonian Center for Astrophysics, 60 Garden Street, Cambridge, MA 02138, USA}

\author[0000-0001-6459-0669]{K. M. Menten} 
\affiliation{Max-Planck Institut f{\"u}r Radioastronomie, Auf dem H{\"u}gel 69, D-53121 Bonn, Germany}




\begin{abstract}
Rotational transitions in vibrationally excited AlO and TiO --- two possible precursors of dust --- were observed
in the 300~GHz range (1~mm wavelength) towards the oxygen rich AGB stars R~Dor and IK~Tau with ALMA, and vibrationally excited AlO was observed towards the red supergiant VY~CMa with the SMA.
The $J=11 \to 10$ transition of TiO  in the $v=1~{\rm{and}}~2$ levels, and the $N = 9 \to 8$ transition in the $v=2$ level of AlO were identified towards R~Dor; {the $J=11 \to 10$ line of}{} TiO was identified in the $v=1$ level towards IK~Tau; and two transitions in the 
$v=1~{\rm{and}}~2$ levels of AlO were identified towards VY~CMa.
{The newly-derived}{} high vibrational temperature of TiO and AlO in R~Dor of $1800 \pm 200$~K, and prior measurements of the angular extent confirm that the majority of the emission is from a region within $\lesssim$2\,$R_{\star}$ of the central star.
A full radiative transfer analysis of AlO in R~Dor yielded a fractional abundance of {$\sim$3\%}{} of the solar abundance of Al. From a similar analysis of TiO a fractional abundance of $\sim78$\% of the solar abundance of Ti was found. The observations provide indirect evidence that
TiO is present in a rotating disk close to the star.
Further observations in the ground and excited vibrational levels are needed to determine whether AlO, TiO, and TiO$_2$ are seeds 
of the Al$_2$O$_3$ dust in R~Dor, and perhaps in the gravitationally bound dust shells in other AGB stars with low mass loss rates.  
\end{abstract}



\section{Introduction} \label{sec:intro}

Most low and intermediate mass stars ($\sim 0.8$--$8~\msol$) pass through the asymptotic giant branch (AGB) phase after leaving the main sequence and exhausting the hydrogen and helium in their cores. High-mass stars ($\gtrsim 8~\msol$), however, pass through the similarly cool red super giant (RSG) phase while fusing helium. Both AGB and RSG stars have high mass-loss rates and produce a significant amount of dust in their molecule rich stellar winds. 
For a comprehensive review of AGB and RSG stars, see \citet{Hofner2018} 
and \citet{Levesque2017}.

The precise formation pathways of dust in AGB and RSG stars are not yet known, but in the past few years astronomers have begun 
observing rotational lines of a few small metal bearing molecules which might be the precursors of the dust.
The high sensitivity and angular resolution of interferometers such as the Atacama Large Millimeter/submillimeter Array (ALMA) 
have allowed astronomers to compare the abundance distributions of small gaseous molecules with the infrared emission from the dust, and to consider the possible connection between the gaseous molecules 
and the dust formation and growth process within a few stellar radii ($R_{\star}$) of the central star \citep{Kaminski2019}.

{Optical and near-infrared} polarimetric images of the dust do not directly reveal the composition and processes by which the dust is formed \citep{Khouri2016,Ohnaka2016}, but parallel studies of the abundance distributions of the gas phase molecules in the dust 
forming region might help unravel the physicochemical processes that occur in the phase change from small molecules 
to gas phase clusters and ultimately tiny dust grains.
\citet{Karovicova2013} used infrared interferometric observations from VLTI/MIDI (the MID-infrared Interferometric instrument on the VLT interferometer) to show that two oxygen rich Mira stars with 
low mass loss rates (S~Ori and R~Cnc) are surrounded by an {amorphous} Al$_2$O$_3$ shell at $\sim$2 stellar radii; while an AGB star 
with a high mass loss rate (GX~Mon) is surrounded by an Al$_2$O$_3$ dust {shell} at around 2 stellar radii and a silicate layer around 
4 stellar radii. 
This led \citeauthor{Karovicova2013} to hypothesise that stars with low mass loss rates primarily form dust whose 
spectral properties match Al$_2$O$_3$, and stars with higher mass loss rate form dust with the spectral properties of warm silicates. 

In oxygen rich AGB and RSG stars (the focus of the work here), the 
most accessible metal oxides and hydroxides in the millimeter band are SiO, AlO, AlOH, TiO, and TiO$_2$.
SiO has been extensively studied since the early days of millimetre astronomy \citep[see for example][]{Gonzalez-Delgado2003}, and several studies have been devoted to AlO,
owing to its relatively intense lines in the millimeter band 
and the possible correlation with the infrared emission from alumina (Al$_2$O$_3$) in the dust forming region \citep[e.g.][]{Tenenbaum2009,Kaminski2016,Decin2017}.  
But at present it is not yet known for certain which metal oxides and hydroxides are the initial building blocks of the identified dust species.

Rotational lines of AlO were first detected in the red supergiant VY~CMa  with the Arizona Radio Observatory (ARO) 10~m antenna by 
\citet{Tenenbaum2009}; and subsequently observed in an interferometric spectral line survey of VY~CMa with the Submillimeter Array (SMA) \citep{Kaminski2013b}; 
and  in Mira  \citep[$o$ Cet,][]{Kaminski2016}, W~Hya \citep{Takigawa2017}, and R~Dor and IK~Tau with ALMA \citep{Decin2017}.

A major contribution toward elucidating the connection of AlO and the dust is the comprehensive study of the Al content 
in R~Dor by \citet{Decin2017}. They concluded that the principal small gaseous precursors of aluminum grains 
(AlO, AlOH, and AlCl) are: 
(1) observed both in and beyond the dust formation region, 
(2) account for only a small fraction of the aluminum budget, and  
(3) condensation of small gaseous aluminum bearing molecules is not 100\% efficient.

Parallel to the work on AlO, rotational lines of TiO and TiO$_2$ were observed with low peak flux in VY~CMa in the same 
spectral line survey with the SMA  \citep{Kaminski2013b,Kaminski2013c}.
Subsequently TiO$_2$ was observed with much higher sensitivity and angular resolution in VY~CMa with ALMA  
by \citet{De-Beck2015}, who concluded TiO$_2$ is not a tracer for ``low grain formation efficiency'' in the complex source 
VY~CMa, because the emission extends well beyond the dust condensation radius. 

Both TiO and TiO$_2$ were also observed with ALMA in three nearby AGB stars:  
Mira~A \citep[the AGB component of a binary,][]{Kaminski2017}, and R~Dor and IK~Tau  \citep{Decin2018}.
\citet{Kaminski2017} concluded on the basis of the observations of TiO and TiO$_2$ 
and indirect chemical arguments, that it is unlikely that substantial amounts of Ti are locked up in solids in Mira~A because: 
(1) the abundance of gaseous titanium species is very high;
(2) a gas phase chemical model is in general agreement with the measured abundances of TiO and TiO$_2$ in the 
millimeter band \citep{Gobrecht2016}; and
(3) TiO$_2$ emission extends $5.5\,R_{\star}$ from Mira~A, rather than close to the star where the temperature 
is higher and inorganic dust is expected to form.
However \citet{Khouri2018} estimated 80\% of the expected Ti in Mira~A is either in the atomic form or in the dust grains 
on the basis of:
(1) a full radiative transfer analysis of CO close to the star and a radiative transfer analysis of dust; 
(2) an LTE analysis of TiO$_2$ in the ground vibrational state; and
(3) the relative abundance of TiO and TiO$_2$ derived by \citet{Kaminski2017}.


There is a general consensus that additional observations with ALMA of AlO, AlOH, TiO, TiO$_2$, and other metal 
oxides and hydroxides are needed to establish more stringent constraints on the dynamical chemical models 
of  the inner winds of oxygen rich AGB and RSG stars.
Conclusive evidence for a strong spatial correlation with the onset of the dust condensation in the inner wind of 
oxygen rich stars can only be made in a credible way if accurate abundance structures of AlO, TiO, TiO$_2$, etc
are retrieved from rotational spectra observed at high sensitivity and angular resolution --- i.e., astronomers 
cannot simply rely on the apparent correlation of millimeter-wave rotational line emission and IR dust emission close 
to the central star.
As previous observations of AlO in the ground vibrational state have shown \citep{Kaminski2016,Decin2017}, 
gaseous AlO and AlOH are observed beyond the dust condensation zone, but these two principal small 
aluminum bearing molecules and AlCl only consume $\sim$2\% of the total Al budget.
Hence, there is ample room left for other Al bearing dust species (or larger gas-phase clusters) to form, and the 
same might hold for TiO and TiO$_2$.
Instead what is needed is a proper assessment of the total Ti consumed in the gas phase species, and whether other
explanations --- e.g., unresolved complex density structures close to the central star --- can be ruled out.


With the aim of advancing the observations of the metal oxides we examined existing interferometric spectral line surveys of two AGB stars (R~Dor and IK~Tau) 
and the RSG star VY~CMa for evidence of vibrationally excited AlO and TiO (Sect.~\ref{sec:Astro}).
However, owing to subtleties in the rotational spectrum of vibrationally excited AlO, and the sparsity of direct laboratory 
measurements of rotational transitions in vibrationally excited TiO, we first reviewed the pertinent molecular physics 
of these two open shell molecules (Sect.~\ref{sec:MolPhys}). 
The identification of millimeter-wave rotational lines of vibrationally excited AlO in three stars (VY~CMa, R~Dor, and IK~Tau), 
and vibrationally excited TiO in two of these (R~Dor, and IK~Tau) is described in Sect.~\ref{sec:Astro}. This is followed by
a full radiative transfer analysis of the rotational lines of AlO and TiO in R~Dor (Sect.~\ref{sec:RadTran}), from which
abundance distributions were derived from observational data
of AlO   
(Sects.~\ref{aloresults}) and 
TiO (Sect.~\ref{tioresults}). 
In Sect.~\ref{sec:discussion} we critically considered the evidence for whether TiO (and by inference TiO$_2$) are seeds of the 
Al$_2$O$_3$ dust in the gravitationally bound dust shell (GBDS) of R~Dor.


In the work here we have adhered to the description of R~Dor in the recent papers by \citet{Khouri2016,Van-de-Sande2018a}, and \citet{Decin2017,Decin2018}.
The semiregular (SR) variable oxygen rich star R~Dor is the closest AGB star to our Sun {\citep[59~pc,][]{Knapp2003}}.
It has a very low mass loss rate {\citep[$1.6\times 10^{-7}\,\spy$, ][]{Maercker2016}}, the stellar diameter is large \citep[54~mas,][]{Norris2012}, it is molecule rich, 
and there is a 27~GHz wide spectral line survey in Band~7 with ALMA \citep{Decin2018} in which 22 molecules and isotopologues have been 
identified. 
 It has been shown from polarized dust emission, 
 at high angular resolution in the {optical and} near IR, that there 
is a high density dust envelope --- or gravitationally bound dust shell (GBDS) --- in R~Dor that is close to the star
\citep[][and references therein]{Khouri2016}. {Modelling such a shell is needed to account for the infrared emission from amorphous Al$_2$O$_3$ and the fractions of scattered light seen towards low mass-loss rate stars such as R~Dor and W~Hya \citep{Khouri2015}.}
The GBDS around R~Dor is located within $1.7 - 1.9\,R_{\star}$ {\citep{Khouri2016}} and 
 is thought to consist of {amorphous} Al$_2$O$_3$ grains or Fe-free silicates with a preference towards Al$_2$O$_3$,
because Al$_2$O$_3$ can condense at higher temperatures. 
The stellar wind is launched from the outer edge of the GBDS \citep[][and references therein]{Van-de-Sande2018a}.
\noindent The focus here is on the inner wind region within {300~mas} of the central star that includes most of the AlO and
TiO emission \citep{Decin2017,Decin2018}, and especially the region within the dust condensation radius at
$\lesssim$\,(2-3)\,$R_{\star}$ of the central star \citep{Van-de-Sande2018a}.

\section{Molecular physics of vibrationally excited A\lowercase{l}O and T\lowercase{i}O\label{sec:MolPhys}}

\subsection{AlO\label{sec:PhysAlO}}

The millimeter-wave rotational spectra of AlO were directly measured in the laboratory in the $v=1$ and $v=2$ levels \citep{Goto1994}
{and are tabulated in \citet{Launila2009} and CDMS\footnote{{The emphasis in this paper is on the new astronomical observations 
of the rotational transitions in vibrationally excited AlO, but a comprehensive list of the laboratory measurements including the ground vibrational
state can be found on the CDMS website at \url{https://cdms.astro.uni-koeln.de/classic/entries/}.}}.}
The spectroscopic constants in the higher vibrational levels ($v = 3, \cdots, 7$) of AlO were derived from rotationally resolved spectra 
of the two lowest electronic transitions \citep[$A^2\Pi \rightarrow X^2\Sigma$ and $B^2\Sigma \rightarrow X^2\Sigma$;][]{Launila2011}.
Rotational transitions in the higher vibrational levels with $v \ge 3$ calculated with the constants of \citet{Launila2011}
should be sufficiently accurate for unambiguous identification in the frequency band in which lines of ground state 
AlO have been observed.

The rotational  lines of AlO in the ground vibrational state are especially wide compared with other molecules such as CO and SiO, whose lines are primarily broadened by the Doppler effect, owing to the expanding envelopes in which they are observed. 
The large observed line widths of AlO are a consequence of the molecular physics: the nuclear spin ($I$) of $^{27}$Al is 5/2; 
the Fermi contact {{magnetic hyperfine}} constant ($b_F$) {{is especially large}};
and the spin-rotation interaction ($\gamma$) in the $X^2\Sigma$ electronic ground state causes the hyperfine structure to appear as two partially overlapping patterns of 5 to 7 hyperfine split components, {{whose centroids are split by approximately $\gamma$}}. 


In most diatomic molecules $\gamma$ changes very little in successive vibrational levels. 
However, owing to a second order interaction of the $X^2\Sigma$ electronic ground state of AlO with the low lying $A^2\Pi$ 
excited state \citep[7775~K above ground,][]{Ito1994}, 
{{the magnitude of $\gamma$ changes from one vibrational level to the next}} 
such that in the ground ($v=0$) vibrational state 
{{\citep[$\gamma= 51.66$~MHz,][]{Goto1994} $\gamma$ is about 3 times larger than in the $v=1$ level 
\citep[$\gamma= 15.95$~MHz,][]{Goto1994}; 
$\gamma$ changes sign between the $v=1$ and $v=2$ levels,\footnote{{{For the rotational transitions of AlO discussed here, 
the theoretical relative intensities of the two spin rotation components are nearly the same:  1.19 for the $N = 6 - 5$ transition and 1.12 for the $N=9-8$ transition, where the $J = N + 1/2$ component is more intense and is at higher frequency if $\gamma$ is positive.}}}
and the magnitude is 1.6 times larger in the $v=2$ than the $v=1$ level \citep[$\gamma= -31.467$~MHz,][]{Goto1994}; 
and the magnitude of $\gamma$ is 3.3~times larger in the $v=3$ level than the $v=2$ level \citep[$\gamma= -103.43$~MHz,][]{Launila2011}.
Therefore the rotational  lines in the $v=1$ level are much narrower than in ground state, but lines in the $v=3$ level}} are 
{predicted to be} very broad and might be difficult to detect.
Because the widths of the rotational lines of AlO vary so markedly from one vibrational level to the next, they provide an independent 
spectroscopic confirmation of the assignments.

\subsection{TiO}
\label{sec:PhysTiO}

\begin{figure*}
\centering
\includegraphics[height=10.6cm]{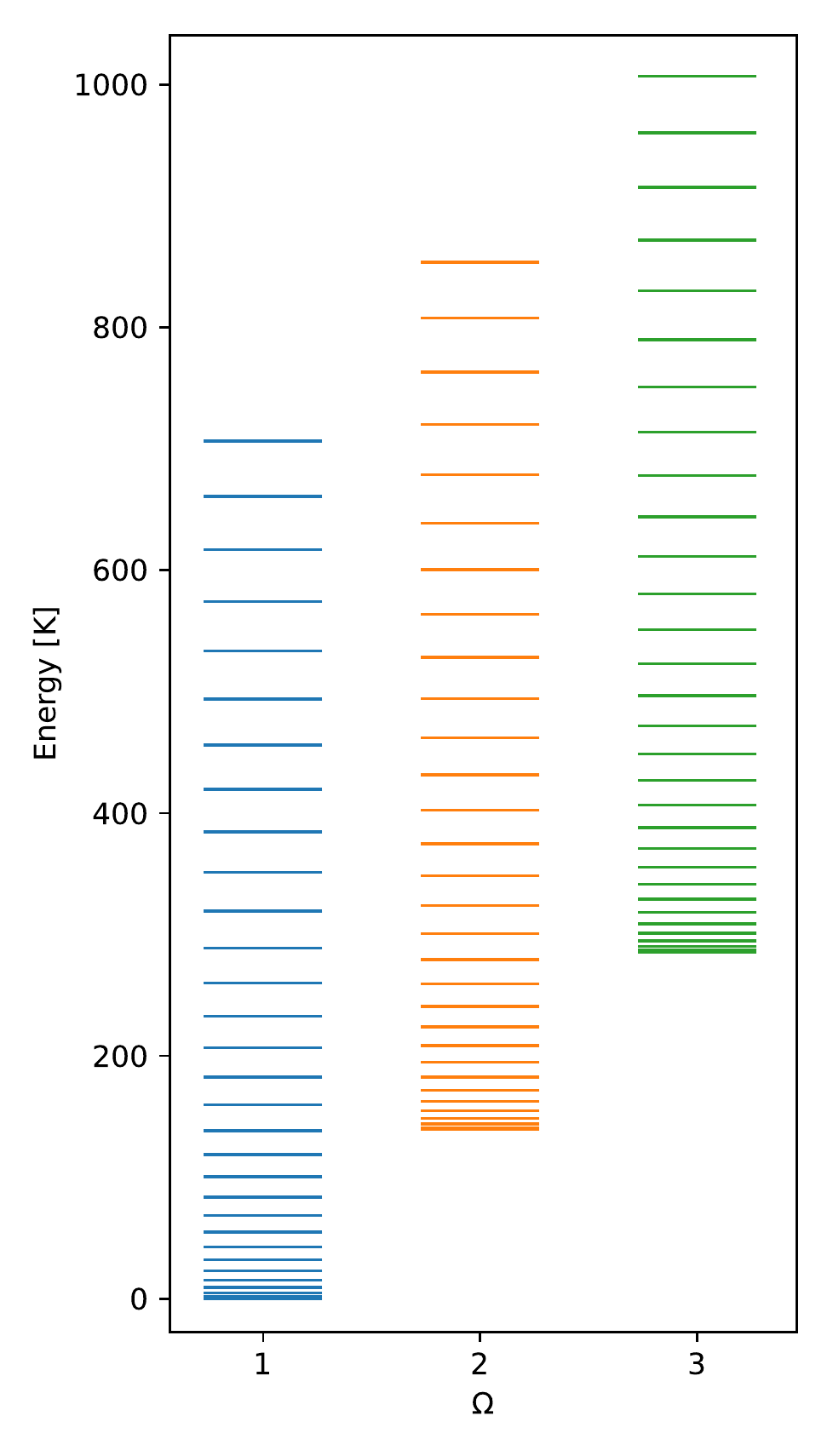}
\includegraphics[height=10.6cm]{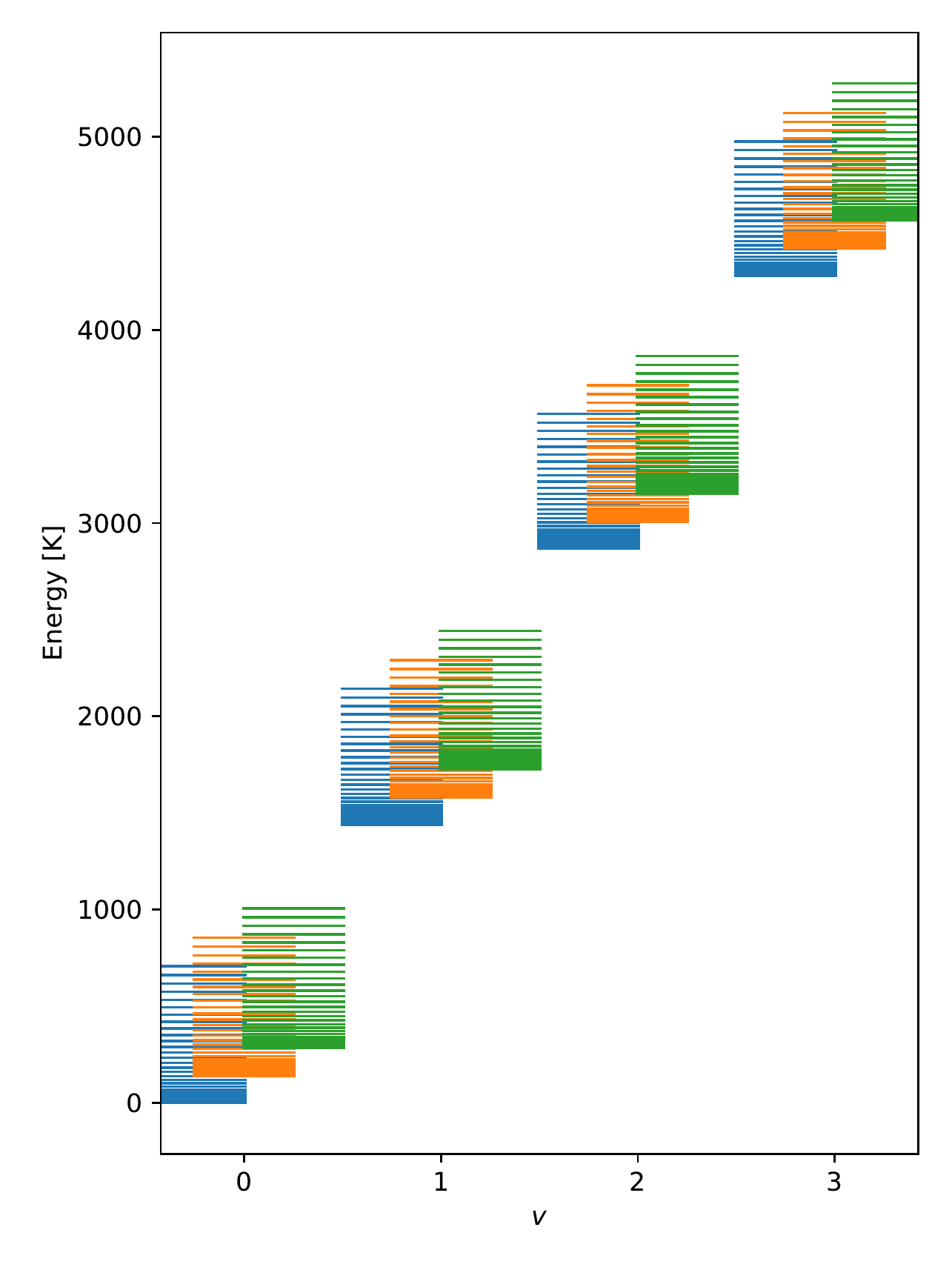}
\caption{Energy level diagram of TiO in the $X^3\Delta_{\Omega}$ ground electronic state 
calculated with effective spectroscopic constants \citep[see Table~2 in][]{Ram1999}.
\textit{Left:} The rotational levels with $J \le 30$ in the ground ($v=0$) vibrational state of the three fine structure components ($\Omega = 1, 2,~{\rm{and}}~3$). 
\textit{Right:} Rotational levels with $J \le 30$ in the ground and three lowest excited vibrational levels
with the same color coding for $\Omega$ as in the panel on the left.\\} 
\label{TiO_ELDrevised}
\end{figure*}

The pure rotational spectrum  of TiO in the ground $X^3\Delta$ electronic state was measured directly in the ground vibrational state \citep{Namiki1998,Steimle2003,Breier2019}, and recently two transitions in the first excited vibrational level were also measured by \citep{Breier2019}.
{{\citet{Namiki1998} measured 12 rotational transitions of TiO between the $J = 6~{\rm{and}}~14$ levels in the $220 - 460$~GHz range in a large absorption cell, and the two lowest transitions at 63 and 95~GHz by microwave optical double resonance; from these \citeauthor{Namiki1998} 
derived the leading spectroscopic constants of TiO  in the $X^3\Delta_i$ electronic ground state.
The same group subsequently measured the permanent electric dipole moment ($\mu$) of the $X^3\Delta$ state in the optical Stark spectrum
of three electronic transitions \citep{Steimle2003}.
More recently \citet{Breier2019} accurately measured 14 rotational transitions of TiO between 253 and 384~GHz with a $1 \sigma$ measurement  
uncertainty of 50~kHz in the ground vibrational state, and two transitions in the $v=1$ level of the $^3\Delta_2$ spin-orbit component in absorption
(at 286.5 and 318.3~GHz with an uncertainty of 100~kHz)  in a supersonic molecular beam.
}}

The most comprehensive spectroscopic analysis of the rotational spectrum of TiO was by \citet{Ram1999}, who derived the rotational and 
fine structure constants in the four lowest excited vibrational levels ($v= 1, \cdots, 4$)  from a combined analysis of rotationally resolved spectra 
of the $A^3\Phi \to X^3\Delta$ electronic transition near $1~\micron$ in the laboratory and in a sunspot, and of the direct measurements in the 
ground vibrational state.
The TiO molecule in its $X^3\Delta_{\Omega}$ ground electronic state  is a good Hund's case $(a)$ molecule with electron spin angular momentum  
($\Sigma$) of 1; orbital angular momentum ($\Lambda$) of 2; $\Omega = \vert { \Lambda + \Sigma} \vert$, where, by vector addition, 
$\Omega$ has the possible values of 1, 2, and 3; and the rotational levels have integral values with $J \ge \Omega$. 
Because the spin-orbit constant ($A$) is positive, the $X^3\Delta_1$ spin orbit component is lowest in energy followed by the $X^3\Delta_2$, and $X^3\Delta_3$ components with relative energies of about 140~K between components 
(see Fig.~\ref{TiO_ELDrevised}).
The three spin components are metastable, because $A$ is much greater than the rotational constant $B$, 
and as a result the line strengths for cross ladder ($\Delta \Omega = \pm 1$) transitions are very small.

\section{Astronomical Observations\label{sec:Astro}}


Many unidentified (U) lines have been observed in spectral line surveys of AGB stars, suggesting the identification of new species 
might be within reach. 
When we considered possible carriers of the U lines, for example towards the oxygen rich stars VY~CMa, R~Dor and IK~Tau \citep{Kaminski2013c,Decin2018}, we realized rotational lines of vibrationally excited AlO and TiO should be readily identifiable for three reasons. 
First, the lowest vibrational levels of both molecules will be populated in the inner wind, because the vibrational frequencies of 
AlO (1389~K) and TiO (1438~K) are comparable to the kinetic temperature of the gas ($T_{\rm{gas}}$) within a few stellar radii 
of the star { (see for example Eq.~\ref{Tgas} in Sect.~\ref{sec:stelparams}).}
Second, a wealth of supporting laboratory spectroscopy for both molecules has been published {{as summarized in Sects.~\ref{sec:PhysAlO}
and \ref{sec:PhysTiO}}}.
And finally, the peak fluxes of the ground state lines in sensitive interferometric spectra are high enough so  
lines in the lowest few excited vibrational levels should be observable in existing surveys.

\subsection{VY~CMa}
\label{sec:VYCMa}

\begin{figure*}
\centering
\includegraphics[width=0.8\textwidth]{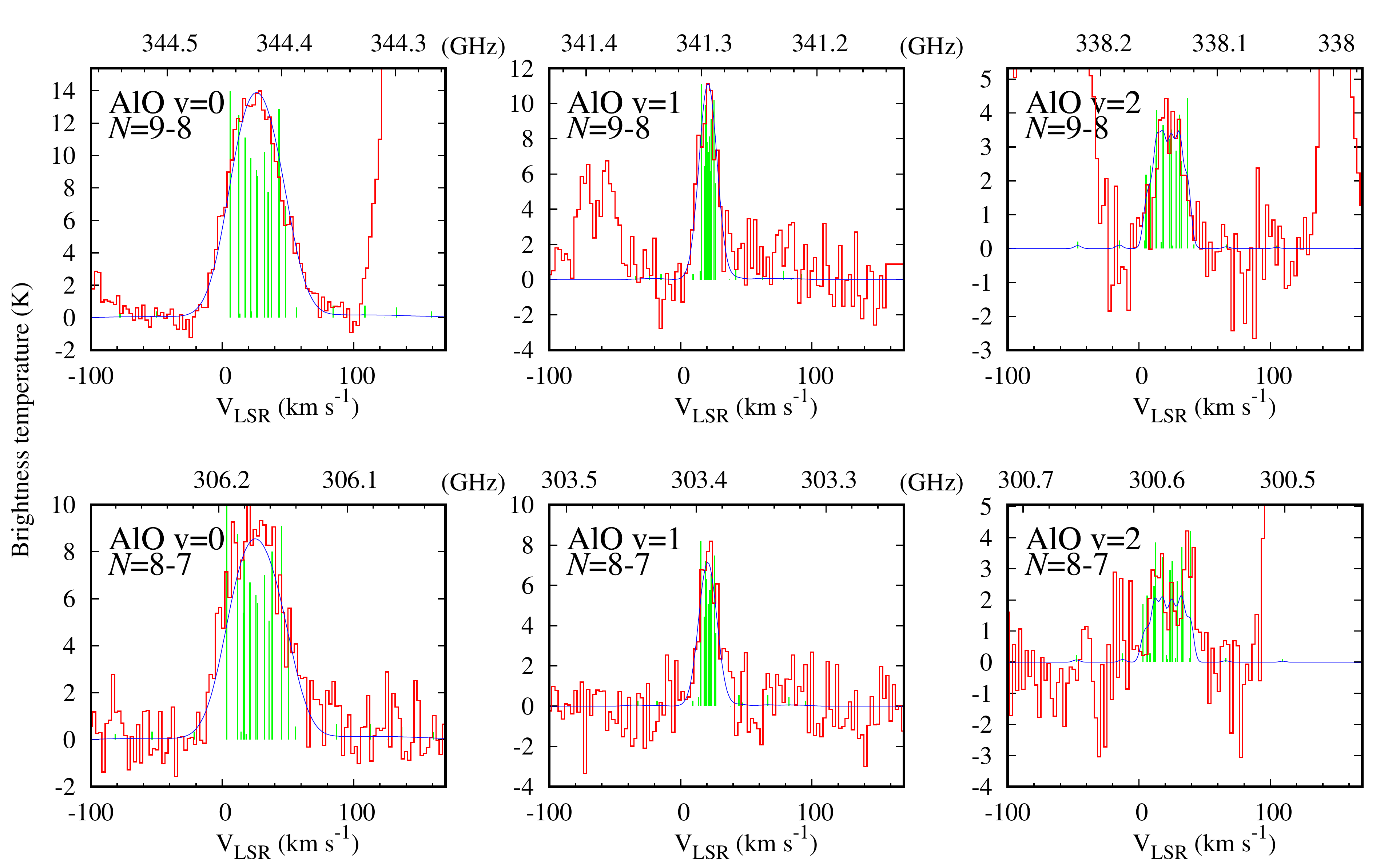}
\caption{ \label{fig:VYCMa} 
Two successive rotational transitions  of AlO observed with a $1^{\prime\prime}$ beam in VY~CMa with the SMA 
\citep{Kaminski2013c}. 
The transition quantum numbers are indicated in the top left of each box.
Data are plotted as red histograms, the unresolved fine and hyperfine structure is indicated by the vertical lines in green, 
and theoretical Gaussian profiles from least squares fits to the observed spectra are plotted in blue. {In the $v=2$ level the full composite synthetic profile with hyperfine structure is plotted, because the spectra were too noisy to 
analyze with a theoretical Gaussian profile.}
 }
\end{figure*}

\begin{deluxetable*}{cccccccchcc}
\centering
\tablecaption{AlO in VY~CMa Observed with the SMA\label{tab:VYCMa}}
\tablehead{
\colhead{$N^{\prime} \to N$}   &  \colhead{$v$}
& \colhead{$\nu_{\rm{calc}}$\tablenotemark{a}}              & \colhead{$\nu_{\rm{obs}}$\tablenotemark{b}}      & 
\colhead{Peak Flux} 
& \colhead{$\Delta \upsilon$}                         & 
\colhead{$\upsilon_{\rm{LSR}}$\tablenotemark{d}}    
& \colhead{Beam}                        & {}                                                 & \colhead{Int. Flux}  & \colhead{Line width}                                
\\
{}&{}&{(MHz)}                                 & {(MHz)}                      
& {(mJy)}
& {(km s$^{-1}$)}             
& {(km s$^{-1}$)}                    & {(\arcsec)}                      & {}                                                    & {(Jy km s$^{-1}$)}                       
& {ratio} 
}
\startdata
$8 \to 7$                            &     0  
& 306,197.319              &    306,172.3   $\pm$ 0.8                 &   580             
&   53.8                  $\pm$     2.0              & 24.5
$\pm$ 0.8                               &      0.91           &                         &     32.4     $\pm$ 1.6  &  1.0    \\ 
                                        &     1  
& 303,412.064              &    303,390.0   $\pm$   0.6                 &   477             
&   17.5                          $\pm$     1.5                &   21.8
$\pm$ 0.6                              &      0.91            &                          &   8.9      $\pm$  0.9    & $3.1 \pm 0.3$   \\ 
                                        &     2 
& 300,611.392              &   \phantom{$^c$}300,591.0   $\pm$   3.0\tablenotemark{c}             &   340             
&   41.0                           $\pm$      7.2             &      20.3
$\pm$ 3.0                               &      0.92           &                       &   7.7        $\pm$     1.6   &  $1.3 \pm 0.2 $  \\ 
$9 \to 8$                            &     0  
& 344,451.721              &   344,423.4    $\pm$   0.3                &   1040             
&   45.8                           $\pm$     0.7              &      24.6
$\pm$ 0.3                               &     0.88            &                         &  50.8      $\pm$  1.0      &  1.0    \\ 
                                        &     1  
& 341,318.153              &    341,294.6   $\pm$   0.6                &   661             
&   12.6                           $\pm$    1.1               &    20.7
$\pm$ 0.5                               &     0.82            &                         &  12.5       $\pm$  1.1     &  $3.6 \pm 0.3$    \\ 
                                        &     2 
& 338,167.232              &    \phantom{$^c$}338,141.0   $\pm$   1.0\tablenotemark{c}              &   242             
&   24.2                           $\pm$    2.1              &     23.3
$\pm$ 0.9                               &    0.81            &                        &   6.2         $\pm$  0.7     &  $1.9 \pm 0.2$   \\ 
\enddata
\tablenotetext{a}{Rest frequencies of the rotational transitions calculated in the absence of fine and hyperfine structure --- see text for details.}
\tablenotetext{b}{{Measured LSR frequency.}}
\tablenotetext{c}{{Frequency estimated for lines with low signal-to-noise.}}
\tablenotetext{d}{{LSR velocity calculated based on the difference between the rest frequency and the measured LSR frequency of each line.}}
\end{deluxetable*}

We began by searching for vibrationally excited AlO in the spectral line survey of VY~CMa in the $279.5 - 355$~GHz range 
observed with the eight element SMA at a modest sensitivity and angular resolution \citep[$\sim$1$^{\prime\prime}$;][]{Kaminski2013c}. 
The initial identifications of vibrationally excited AlO were made on the basis of the calculated rotational frequencies 
in the three lowest excited vibrational levels ($v \le 3$) in the absence of fine and hyperfine structure 
(i.e., with $\gamma, b_F$, {{the dipole-dipole magnetic hyperfine constant} $c$, and {{the electric quadrupole coupling constant}} $eQq$ constrained to zero).
{Although the frequencies of the rotational transitions of AlO calculated with only two spectroscopic constants 
({{the rotational constant}} $B$ and {{centrifugal distortion constant}} $D$) are approximate, 
the centroids of the synthetic line profiles reproduce the astronomical features to within a few MHz and are sufficient 
for establishing the identification of lines of vibrationally excited AlO in the astronomical spectra.}
{{The frequencies and relative intensities of the hyperfine components used to calculate the synthetic profiles of AlO in the 
$v = 0, 1,\,{\rm{and}}~2$ levels are tabulated in \citep{Launila2009}.}}
The term energy for the $v = 1$ excited vibrational level is {{1389~K}} and {{2758~K for the $v = 2$ level}}.

Two successive rotational transitions of AlO in the ground vibrational state ($N = 8 \to 7$ at 306~GHz, and $N=9 \to 8$ at 344~GHz) 
were covered in the SMA survey, and both were also observed in vibrationally excited levels: the $v=1$ {{and possibly the
$v=2$ level}} in the $N=8\to 7$ transition, and $v=1$ and $v=2$ {{levels}} in the $N=9\to 8$ transition (see Fig.~\ref{fig:VYCMa} {and Table \ref{tab:VYCMa}}).
The line parameters for most measurements were derived by fitting Gaussian profiles to the astronomical features.
{All measurements were done on spectra obtained from the central pixel of the SMA data, corresponding to a restoring beam of $\sim1\arcsec$ centred on the continuum peak of VY CMa \citep[for precise beam sizes at different frequencies, see][]{Kaminski2013c}.}
The observed frequencies $\nu_{\rm{obs}}$ {in the ground vibrational state and the $v=1$ level} {and the FWHM line width, $\Delta v$,} were derived from Gaussian profiles that were least squares fit to
the spectra (see Fig.~\ref{fig:VYCMa}).
{The synthetic profiles of the lines in the $v=2$ vibrational state were obtained by representing each hyperfine component with a Gaussian profile whose HWHM was 2.5~\kms.  The relative intensities of the hyperfine components were equivalent to the theoretical relative intensities in \cite{Launila2009}. The variables in the least squares fit of the synthetic profile to the observed profile were: an intensity (scaling) factor of the composite theoretical synthetic profile and the center frequency.}
{{The mean central LSR velocity $\upsilon_\mathrm{LSR}=21.5 \pm 3.3$~km~s$^{-1}$ for the four rotational transitions in the excited vibrational levels of AlO in Table~\ref{tab:VYCMa} is in good agreement with the {stellar velocity \citep[$v_\mathrm{LSR}=22$~\kms,][]{Kaminski2013c} and the LSR velocity of AlO in the ground vibrational state (}$\upsilon_\mathrm{LSR}=25 \pm 2$~km~s$^{-1}$) observed with the SMA in VY~CMa by \citet{Kaminski2013c}, confirming the assignment of the new lines of vibrationally excited AlO.}}
%
As expected, the FWHM of the lines  are $3.3 \pm 0.4$ times smaller in the $v=1$ level,
and $1.6 \pm 0.3$ times smaller in the $v=2$ level than those in the ground vibrational state --- i.e., close to the theoretical 
predictions discussed in Sect.~\ref{sec:PhysAlO}, thereby providing further confirmation of the line assignments. 
The rotational lines in the $v=3$ vibrational level were not observed because the predicted peak flux is 
comparable to the peak noise in the spectra shown in Fig.~\ref{fig:VYCMa}.
The details of the detected lines are given in Table~\ref{tab:VYCMa}.

\subsection{R~Dor and IK~Tau}\label{almadatadesc}


\begin{deluxetable*}{lccccchhcccc}
\tablecaption{Details of ALMA observations of AlO and TiO towards R~Dor and IK Tau.}
\label{tab:almaobs}
\tablehead{\colhead{Transition}&Freq    & Proposal      & Date(s)   & ToS\tablenotemark{a} &PWV\tablenotemark{b}  &Bandpass/     &Phsref    & Synth. beam\tablenotemark{c}    &Vel. res. & rms\tablenotemark{d} &MRS\tablenotemark{e}\\
          & (GHz)  &               & (yyyymmdd)&(min)& (mm)&Flux Scale cal&          &(mas,mas,deg) &(km s$^{-1}$)&(mJy)&(arcsec)
}
\startdata
\bf R Dor     &        &               &           &     &     &              &          &              &            &     &   \\
AlO $N=6\to5$ &229.670 &2017.1.00824.S & 20171024-- &192  & 0.6 &J0635-7516,   &J0516-6207&      33, 29, +55        &  0.32          &  0.8   &2.6\\
    	  &	   &		   & 20171026  &     & 	   &J0519-4546,	  &	     &		    &		 &     &   \\
AlO $N=7\to6$ &267.937 &2017.1.00824.S & 20171231-- &165  & 1.2 &J0635-7516,   &J0516-6207&       138,131,$-55$       &   0.273         &   1.0  &5.0\\
    	  &	   &		   & 20180105  &     & 	   &J0522-3627    &	     &		    &		 &     &   \\
AlO $N=8\to7$ &306.197 &2017.A.00012.S & 20181211  & 6.5 & 0.9 &J0635-7516    &J0516-6207&      110, 88, $-45$        &  0.911          & 3.8    & 3.1\\
AlO $N=9\to8$&338.167--&2013.1.00166.S & 20150827-- &25.2 &0.3-- &J0522-3627    &J0506+6109& 165, 127, $+29$ &  0.9--     & 3.5--& 3.1\\
\quad \& TiO lines  &355.623 &               & 20150901  &     &0.7  &J0457-2324    &J0428-6438& 150, 125, $+21$ &  0.8       & 4.5 &   \\
\bf IK Tau    &        &               &           &     &     &              &          &              &            &     &   \\
AlO $N=9\to8$&338.167--&2013.1.00166.S & 20150813-- &10   &0.2-- &J0423-0120     &J0407+0742& 175, 145,$-15$ &  1.7--       & 4.7--& 3.1\\
\quad \& TiO lines  &355.623 &               & 20150828  &     &1.7 &               &          & 165, 135,$-10$ &  1.6       & 5.3 &   \\
\enddata
\tablenotetext{a}{Time on source, interleaving with phase reference
observations every 5--7 min.}
\tablenotetext{b}{Precipitable water vapour at time of observation.}
\tablenotetext{c}{{Listed are the major and minor axes, and the position angle of the synthetic beam.}}
\tablenotetext{d}{The rms is the noise per channel off-source in the line cube.}
\tablenotetext{e}{Maximum recoverable scale; reliable images can only be made at slightly smaller scales.}
\end{deluxetable*}

The data used in this paper were taken from the ALMA projects with
codes and other observing details given in Table \ref{tab:almaobs}; see the ALMA Science Archive for more details. 
The raw data were calibrated using
standard ALMA procedures including application of instrumental
calibration and corrections derived from the bandpass and phase
reference sources.  The flux scale is in most cases accurate to 7\% or
better; however, for AlO $N=6\to5$ the phase reference source flux density
fluctuated by 12\% over the observations, which is probably bona fide
variability but gives an upper bound to the uncertainty for these data.
\cite{Decin2018} gives details of the data reduction for 2013.1.00166.S
and we followed a similar route in all cases. In brief, line-free
channels were identified and used for self-calibration of the strong
continuum emission from the star (and potentially hot dust).
 After applying all calibration to the line data,
spectral cubes were made at the resolutions and sensitivities given in
Table \ref{tab:almaobs}, averaging where necessary to improve surface brightness
sensitivity. 

The continuum was unresolved or barely resolved, with a flux density ranging from 252~mJy in the image covering AlO $N=6\to5$, to 596~mJy covering AlO $N=9\to8$. The relative astrometry between continuum and line in the same data-set is limited by the synthesised beam/signal-to-noise ratio (S/N). This is $<1$~mas for all continuum images. For the moment maps, the relative position uncertainty is $\sim$5~mas for the weakest emission, down to $<1$ mas for the brightest --- less than a pixel in all cases.

We extracted spectra in an aperture of {150~mas radius for AlO and 300~mas for TiO towards} R~Dor and in an aperture of 320~mas {for AlO or 160~mas for TiO} for IK~Tau. 
The peak and integrated fluxes given in Tables \ref{tab:AlOalma} and \ref{tab:TiOalma} were measured from these spectra.
We made zeroth-order moments (integrated flux density) over the channels where the line signal exceeded the
rms. 
For the cases where lines participate in overlaps with lines of other species, the details are discussed
in Sections \ref{sec:AlOalma} and \ref{sec:TiOalma} for AlO and TiO, respectively. Due to the quality and quantity of data, we mostly focus on R~Dor for our analysis.

\subsubsection{AlO}
\label{sec:AlOalma}


\begin{figure*}
\centering
\includegraphics[width=0.327\textwidth]{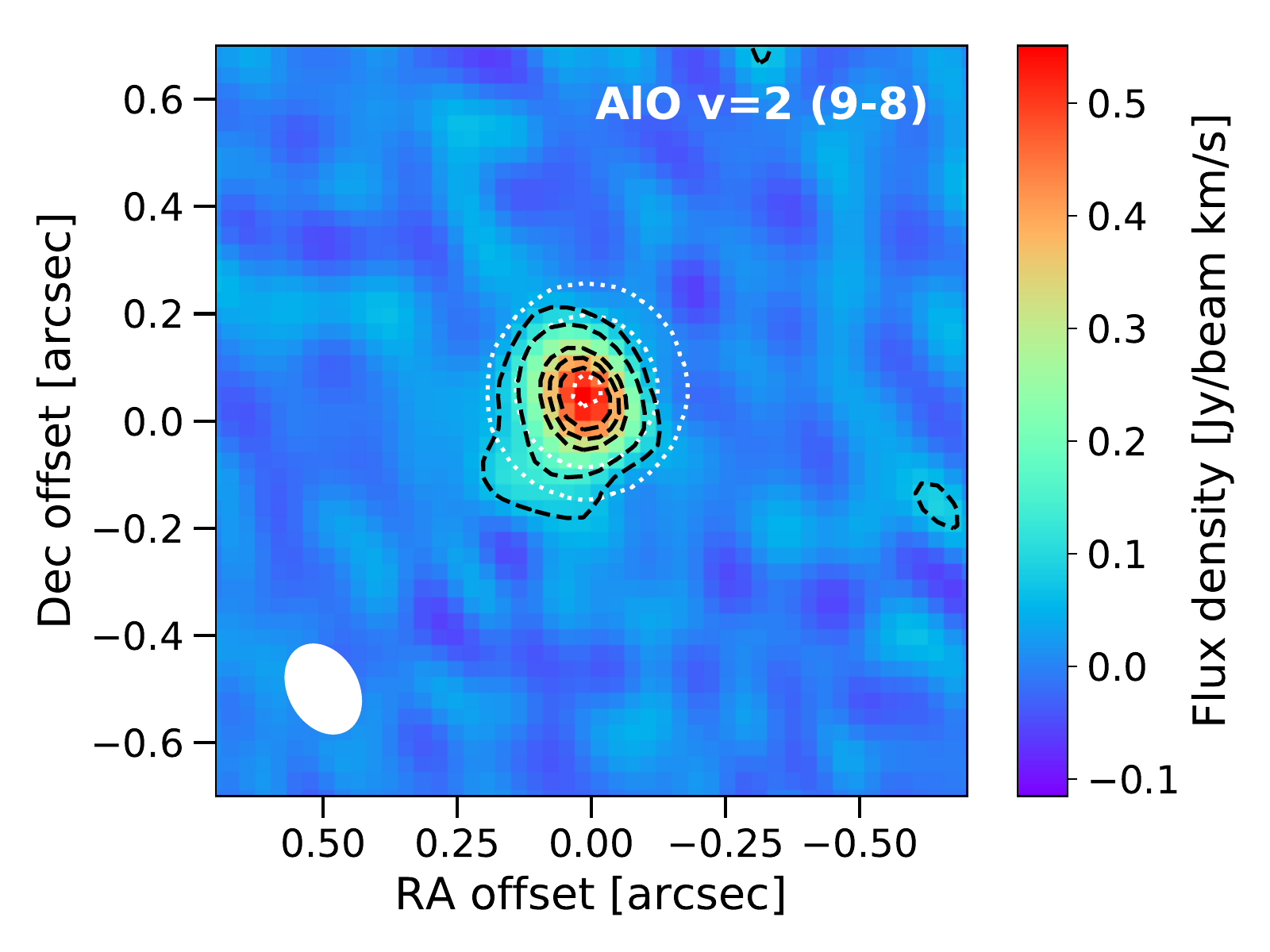}
\includegraphics[width=0.32\textwidth]{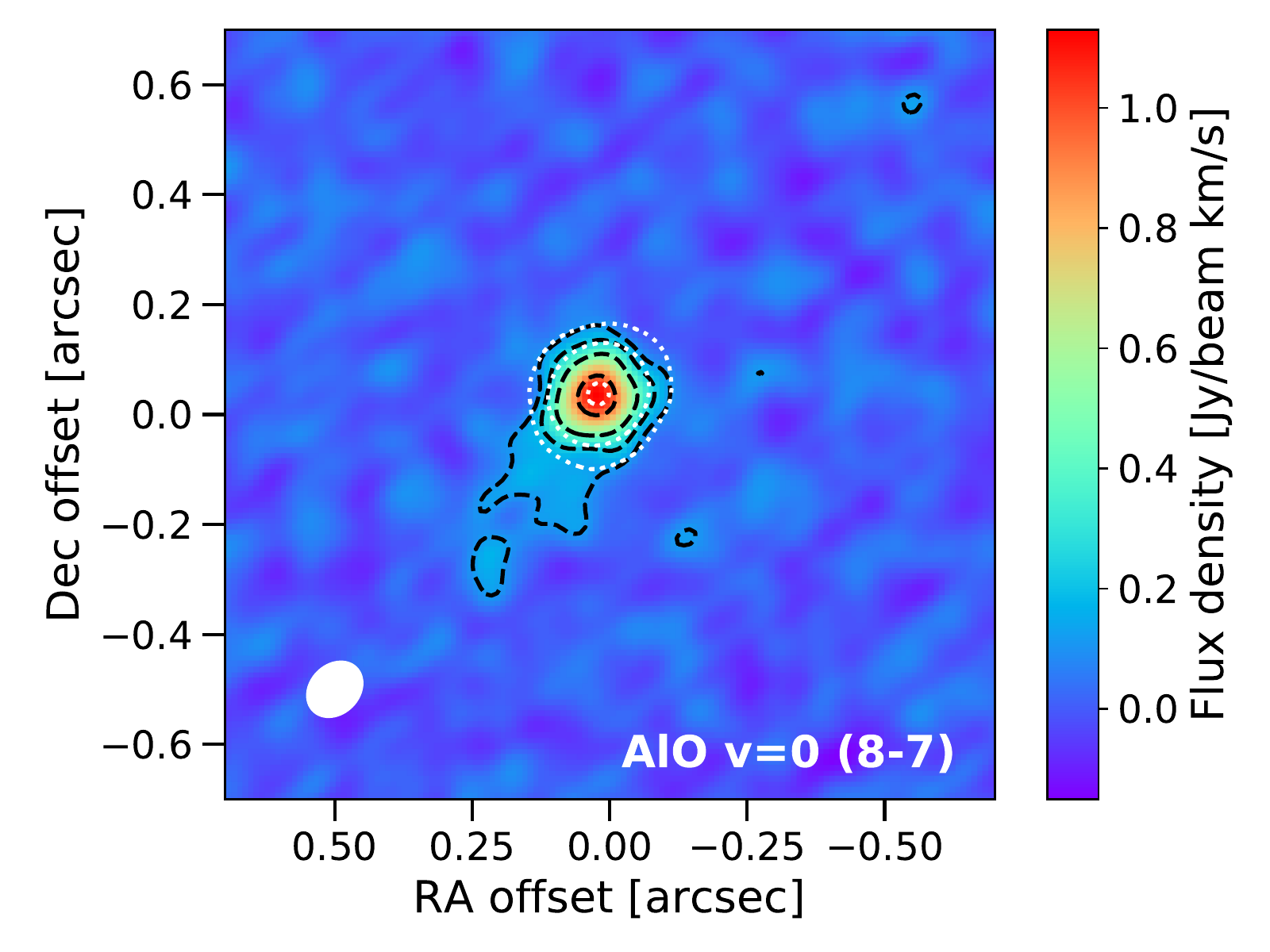}
\includegraphics[width=0.32\textwidth]{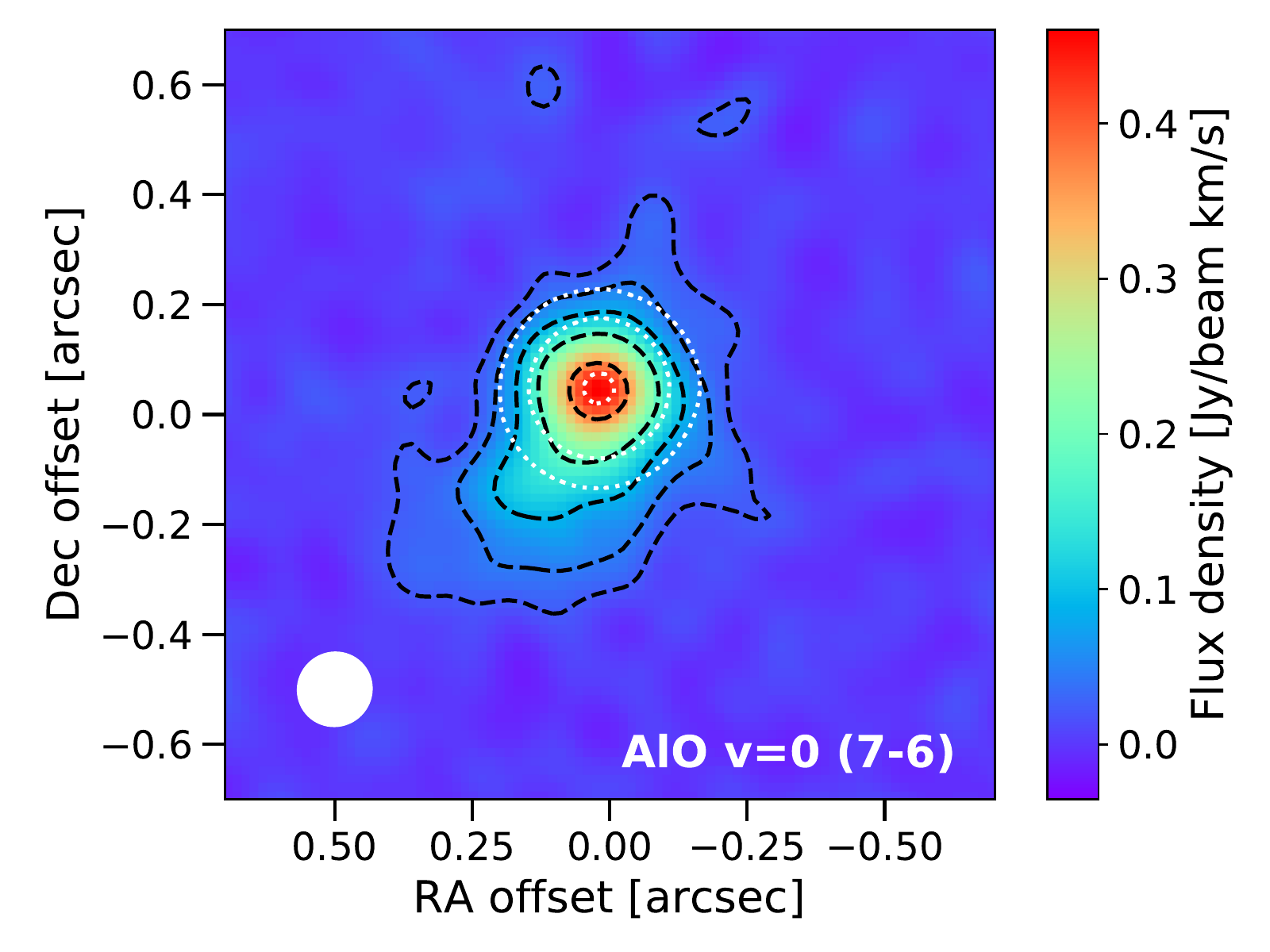}
\caption{\label{fig:AlOmaps} For R~Dor: zeroth moment maps of rotational transitions of AlO in the ground ($X^2\Sigma$) electronic state, observed with ALMA. \textit{Left:} $N=9\to8$ in the $v=2$ vibrational state. \textit{Centre:} $N=8\to7$ in the ground vibrational state. \textit{Right:} $N=7\to6$ in the ground vibrational state.
The black dashed contours indicate flux levels at 3, 6, 12... times the off-source noise rms for AlO.
The synthesized beams are indicated by ellipses in the lower left corner of each panel.
The dotted white contours are at levels of 1, 10, and 90\% of the peak continuum flux and are included to indicate the position of the star.\\
}
\end{figure*}


\begin{deluxetable*}{c c cc c ccc c}
\tablecaption{The observed transitions of AlO towards  R~Dor and IK~Tau.}
\label{tab:AlOalma}
\tablehead{ 
\multicolumn{1}{c}{$v$}       
&  \multicolumn{1}{c}{$N$}   
&  \multicolumn{2}{c}{Frequency (MHz)} 
&  \multicolumn{1}{c}{$\upsilon_\mathrm{LSR}$\tablenotemark{e}}
&  \multicolumn{1}{c}{Peak flux}   
& \colhead{$\Delta \upsilon$}   
& \colhead{Aperture}    
& \colhead{Integrated flux}
\\    
\cline{3-4}             {}  
                                            & {}  
& {Laboratory\tablenotemark{a}}       & { Observed\tablenotemark{b}}      & {(km~s$^{-1}$)}
& {(mJy)}                                               & {(km~s$^{-1}$)}                                    & {(mas)}    
& {(mJy~km~s$^{-1}$)} 
}
\startdata
\multicolumn{5}{c}{}         &   \multicolumn{1}{c}{{\bf R~Dor}}   &\multicolumn{2}{c}{}             \\
0 & $6\to5$ & 229,670.239 & $\cdots$\tablenotemark{c} & $\cdots$ &$\cdots$& $\cdots$&$\cdots$& $\cdots$\\
0 & $7\to6$ & 267,936.560 & $267,932.6\pm5.4$ & 4.4 &$19\pm3$& 73&150& $775\pm23$\\
0 & $8\to7$ & 306,197.319 & $306,191 \pm 12$ & 5.4 & $50\pm6$& 68&150& $1986\pm 90$\\
 0                                      &      $9 \to 8$                            
&  344,451.721                 & $344,445.1 \pm 1.5$                             &                 
5.8
&   $87 \pm 12$             & 51.0                                     &     150                   & $3133 \pm 75$          \\
2                                      &      $9 \to 8$                            
&   338,167.232               &    $338,156 \pm 12$                          &   10.3              
&   $22 \pm 5$                &    29.7      &       150                    &   $612 \pm 14$        \\
\multicolumn{5}{c}{}        &   \multicolumn{1}{c}{{\bf IK~Tau}\tablenotemark{d}}   &\multicolumn{2}{c}{}             \\
0                                      &   $9 \to 8$                               
&  344,451.721                & $\cdots$                           &     $\cdots$            
&   $23$                           & $\cdots$                           &     320                  &   247         \\
2                                      &          $9 \to 8$                        
&   338,167.232               &    $\cdots$                        &    $\cdots$             
&   $\lesssim 10$             &    $\cdots$                       &       320                &   $\cdots$       \\
\enddata
\tablenotetext{a}{Frequencies of the rotational transitions given in the absence of fine and hyperfine structure --- see text for details.}
\tablenotetext{b}{Measured LSR frequency.}
\tablenotetext{c}{The high spatial resolution of this observation resulted in an image dominated by absorption against the star, complicating the interpretation.}
\tablenotetext{d}{ The peak and integrated flux values of the $N=9\to8$, $v=0$ line for IK~Tau were taken from \cite{Decin2017}. All other values were obtained from our present analysis.}
\tablenotetext{e}{{LSR velocity calculated based on the difference between the rest frequency and the observed LSR frequency of each line.}}
\end{deluxetable*}

Following the initial identification of vibrationally excited AlO in the SMA survey of VY~CMa, we found evidence for the 
$N = 9 \to 8$ transition in the $v = 2$ level ($E_u = 2758$~K) with an approximate peak flux of $30 \pm 8$~mJy at the expected frequency and line width (see {Table~\ref{tab:AlOalma})} in a sensitive spectral line survey of R~Dor over the 335--362~GHz frequency range with ALMA \citep{Decin2018}. 
Unfortunately the $N = 9 \to 8$ transition in the $v=1$ level of AlO at 341,318~MHz is within $3-4$~MHz of two transitions 
of SO$_2$ \citep[for details of SO$_2$ in R~Dor see][]{Danilovich2020}, 
and the rotational transition in the $v=3$ level at 334,999~MHz lies just outside of the frequency range of the survey.
The corresponding lines of SO$_2$ are much less prominent in the SMA survey of VY~CMa than in the ALMA survey
of R~Dor, and did not preclude the identification of the $N = 9 \to 8$ transition in the $v=1$ level in VY~CMa. 

In addition to these previously published data, we have obtained some new observations using ALMA, in a search for the $N=6\to5$ through $N=8\to7$ rotational transitions in the ground vibrational state. 
In Table \ref{tab:AlOalma} we list the frequencies of all the observed AlO transitions in the absence of fine and hyperfine structure calculated from 
the rotational constant $B$ and centrifugal distortion constant $D$ derived 
from the measurements tabulated in \citet{Lovas1974}. 
The full set of AlO observed spectra towards R~Dor is plotted with our radiative transfer model results in Fig.~\ref{fig:AlO}.







The approximate location and extent of AlO in R~Dor was determined from zeroth moment and channel maps obtained with ALMA \citep[initially from the spectral line survey of][ which shows the channel maps of AlO $N=9\to8$, $v=0$ towards R~Dor in their Fig.~3]{Decin2017}.
Although the AlO emission mainly coincides with the continuum emission, there is additional emission to the southeast 
that extends out to about 280~mas from the star, as can be seen for the $N=8\to7$ and $N=7\to6$ zeroth moment maps in Fig. \ref{fig:AlOmaps}.
The $N=9\to8$ emission (in both $v=0,2$ vibrational states) is only just resolved. Nevertheless, from the azimuthally averaged flux density versus distance from the star, \cite{Decin2017} determined that approximately 50\% of the AlO emission is within $\sim$3\,$R_{\star}$ of the central star (see their Fig.~4).

\begin{figure}
\centering
\includegraphics[width=\columnwidth]{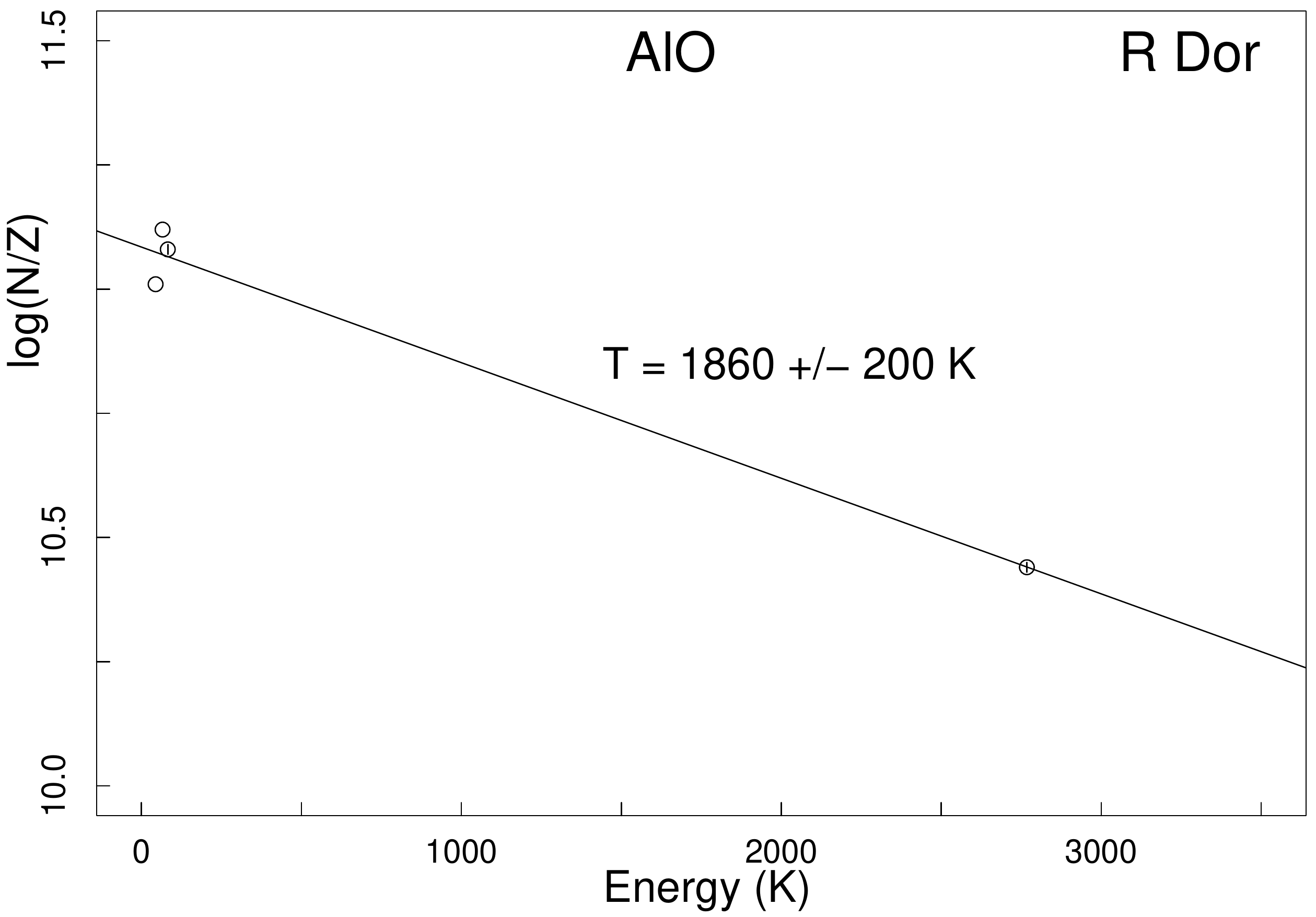}
\caption{Vibrational temperature diagram for AlO in R~Dor derived from three successive rotational transitions in the  
ground vibrational state and one in the second excited ($v=2$) vibrational level.  }
\label{AlO:RDorvib}
\end{figure}


{{Shown in Fig.~\ref{AlO:RDorvib} is a vibrational temperature (VT) diagram for AlO in R~Dor.
The purpose of the VT diagram is to establish: 
(1) whether there is any evidence for anomalies in the emission in the ground and excited vibrational levels;
(2) whether the observed emission can be analyzed in the 1D approximation; and
(3) whether the derived vibrational temperature ($T_{\rm{vib}}$) can inform us about the region where AlO is observed.
The VT diagram was constructed with the integrated flux of the $N = 9 - 8$ transition in the $v =2$ excited vibrational level, and 
three transitions in the ground vibrational state (Fig.~\ref{AlO:RDorvib}),
and the standard expression \citep[see Eq.~A5 in][]{Snyder2005} on the assumption the emission fills the synthesized beam of the telescope.
The partition function ($Z$) includes all the rotational-vibrational levels with all the appropriate degeneracy factors in the comprehensive analysis 
of \citet{Patrascu2015}.
From this we determined the total column density ($N$), but it should be emphasized that a more informative description of the abundance 
of AlO in the inner wind is obtained from the full radiative transfer analysis discussed in detail in Sect.~\ref{sec:RadTran}.
The high observed $T_{\rm{vib}}$  of AlO in R~Dor of $1860 \pm 200$~K implies a fraction of the emission arises within about 
$2\,R_{\star}$ (or $\sim$60~mas) from the central star --- i.e., a fraction of the AlO emission is within the dust formation zone of R~Dor at 
$\lesssim 2\, R_{\star}$ \citep{Khouri2016} --- whose effective temperature ($T_\mathrm{eff}$) is 2400~K \citep{Maercker2016}.}}


Although the AlO emission in the $N = 9 \to 8$ line in the ground vibrational state is very weak in IK~Tau,  \citet{Decin2017} 
succeeded in deriving an abundance distribution, but the spectra in IK~Tau are not analyzed here because no new lines
were observed in the ground vibrational state and the line in the $v=2$ level at 338,167~MHz is exceedingly weak.

\subsubsection{TiO}
\label{sec:TiOalma}


\begin{deluxetable*}{cc c cc c ccc c}
\tablecaption{The $J = 11 \to 10$ transition of $X^3\Delta$ TiO in  R~Dor and IK~Tau}
\label{tab:TiOalma}
\tablehead{
\multicolumn{1}{l}{$\Omega$}   &  \multicolumn{1}{c}{$v$}         
&  \multicolumn{1}{c}{}   
&  \multicolumn{2}{c}{Frequency (MHz)} 
&  \multicolumn{1}{c}{$\upsilon_\mathrm{LSR}$\tablenotemark{d}}
&  \multicolumn{1}{c}{Peak flux}   
& \colhead{$\Delta \upsilon$}   
& \colhead{Aperture}    
& \colhead{Integrated flux}
\\    
\cline{4-5}  
 {}                   & {}                                            & {}   
& {Laboratory}       & { Observed\tablenotemark{a}}      & {(km~s$^{-1}$)}
& {(mJy)}     & {(km~s$^{-1}$)}                    & {(mas)}    
& {(mJy~km~s$^{-1}$)} 
}
 
\startdata
\multicolumn{6}{c}{}        &   \multicolumn{1}{c}{{\bf R~Dor}}   &\multicolumn{2}{c}{}             \\
     1                                 &    0                                     &                                  
&  348,159.8                    & $348,150.75\pm 0.22$    &     7.8            
&   $221 \pm 9$               & $9.4 \pm 0.4$                   &        300                  &   $2,077 \pm 122$         \\
                                        &   1                                      &                                  
&    346,198.2                  &  $346,188.8\pm 0.4$       &      8.1             
&    $91 \pm7$                &  $8.8 \pm 0.9$                  &         300                 &   $804 \pm 103$         \\
                                        &    2                                    &                                  
&   344,224.3                   &    $344,225\pm 3.1$       &      7.5           
&   $44\pm4$                           &       $8.2\pm0.9$                     &          300              &   $377\pm13$          \\
                                        &   3                                     &                                  
&    342,239.0                &    $\cdots$\tablenotemark{b}                        &   $\cdots$              
&   $\cdots$              &    $\cdots$                        &         300               &     $\cdots$       \\
     2                                 &    0                                         &                                  
&   352,157.6                &    $352,148.18\pm 0.25$   &      8.0           
&   $184 \pm 8$            &    $9.4 \pm 0.5$                   &       300                     &   $1,739 \pm 119$      \\
                                         &        1                                    &
&     350,147.5               &  $350,138.9\pm 0.4$       &     7.4            
&   $88 \pm 7$               &    $9.1 \pm 0.9$                 &        300                    &   $799 \pm 101$      \\
                                       &    2                                        &                                  
&   348,125.8                & ($348,112.1 \pm 0.17$)\tablenotemark{c}              &                 $\cdots$
&   ($73 \pm 16$)            &   ($12.9 \pm 0.3$)              &       300                      &    (943)        \\
                                        &   3                                        &                                  
&    346,090.1                &    $\cdots$                         &  $\cdots$               
&    $12 \pm 1$               &    $8.7 \pm 5.2$                           &         300                   &     $\cdots$       \\
     3                                &    0                                        &                                  
&   355,623.3                &   $355,614.38\pm 0.26$   &          7.5       
&     $208 \pm 8$         &   $11.0 \pm 0.5$                  &        300         &    $2,297 \pm 137$         \\
                                       &   1                                         &                                  
&    353,573.4               &   $353,575.2\pm 0.7$       &           8.5      
&     $75 \pm 11$         &      $8.4 \pm 1.4$                 &      300            &    $630  \pm 140$            \\
                                        &    2                                       &                                  
&    351,511.2                &   $351,502.5\pm 1.5$      &              7.5   
&     $51 \pm 14$          &    $7.7 \pm 2.4$                  &      300           &     $391 \pm 127$              \\
                                        &   3                                        &                                 
&   349,436.5                 &   $\cdots$         &      $\cdots$           
&    $14 \pm 1$                &    $17.2 \pm 3.4$                           &         300         &     $\cdots$       \\
\multicolumn{6}{c}{}    &   \multicolumn{1}{c}{{\bf IK~Tau}}   &\multicolumn{2}{c}{}             \\
     1                                &    0                                       &                                  
&  348,159.8                 & $348,159.6 \pm 0.8$        &     32.2            
&   $25 \pm 5$              & $6.6 \pm  1.5$                   &        160                    &   $230$        \\
                                        &   1                                       &                                  
&    346,198.2               &  $\cdots$                            &    $\cdots$               
&   $\cdots$                   & $\cdots$                                             &         160                   &      $\cdots$      \\
     2                                 &    0                                       &                                  
&   352,157.6                &    $352,157.1 \pm 0.8$      &             31.2    
&   $30 \pm 5$            &    $8.0  \pm 1.5$                   &       160                   &   $198$       \\
                                      &        1                                   &
&     350,147.5               &  $350,147.3 \pm 2.5$       &            32.4     
&   $17 \pm 4$               &    $6.0 \pm 5.1$               &        160                  &  139      \\
     3                                &    0                                        &                                  
&   355,623.3                &   $355,625.0\pm4.1$                            &   32.2              
&     $41\pm12$                 &   $6.6\pm5.7$                            &        160                 &   $239$         \\
                                       &   1                                         &                                  
&    353,573.4              &   $353,575.7 \pm 1.6$       &        32.2         
&     $15 \pm 4$         &      $4.2 \pm 3.4$                &      160                     &     75              \\
\enddata
\tablenotetext{a}{ Measured LSR frequency.}
\tablenotetext{b}{   The $20_{1,19} - 19_{2,18}$ transition of $^{34}$SO$_2$ ($E_u = 198$~K) at 342,232~MHz 
and peak flux 93~mJy is within 7~MHz of the transition in the $v = 3$ level of $^3\Delta_1$ TiO.}
\tablenotetext{c}{   The rotational line in the $v=2$ level of $^3\Delta_2$ TiO has a lower observed frequency than expected, and large 
line width, owing to a blend with the $19_{4,16} \to 19_{3, 17}$ transition of $^{34}$SO$_2$. }
\tablenotetext{d}{{LSR velocity calculated based on the difference between the rest frequency and the observed LSR frequency of each line.}}
\end{deluxetable*}


Following the identification of vibrationally excited AlO we turned our attention to TiO. 
The three spin components of the $X^3\Delta_{\Omega}$ electronic ground state (labelled $\Omega = 1,2,3$) were observed 
in the spectral line survey of R~Dor and IK~Tau by \citet{Decin2018}.
Frequencies of the rotational transitions in the ground vibrational state were calculated from constants derived 
from the measurements in \citet{Namiki1998,Breier2019}, and CDMS; 
frequencies in the $v = 1,2,3$ excited vibrational levels were calculated from the spectroscopic constants in \citet{Ram1999}.
The term energies of the four lowest excited vibrational levels ($v = 1,2,3,\,{\rm{and}}\,4$) are 1438~K, 2863~K, 4275~K, 
and 5673~K.

In all, five lines of vibrationally excited TiO were identified towards R~Dor:  three in the $v = 1$ level, and two in the $v = 2$ level {(see Table \ref{tab:TiOalma})}.
The $J =11 \to 10$ transition in the $v=2$ level of $^3\Delta_2$ is uncertain, because it is blended with a line of $^{34}$SO$_2$ 
which is 8~MHz lower in frequency. 
The TiO emission towards R~Dor is not spatially resolved in our observations, indicating that it is less spatially extended than AlO.
Frequencies and other details of these lines are listed in Table~\ref{tab:TiOalma}.
Based on the line frequencies reported in Table \ref{tab:TiOalma}, we find the mean central velocity of the TiO lines towards R~Dor is $\upsilon_{\rm{cent(TiO)}}=7.8 \pm 0.4$~\kms, based on fitting Gaussians to the emission for lines with $v\leq 2$ and excluding the line blended with SO$_2$. This is slightly red-shifted from the $\upsilon_\mathrm{LSR}=7.0\pm0.5$~\kms{} {of R~Dor} {\citep{Danilovich2016}}. A similar small shift {of $\sim1$~\kms{}} was also seen for vibrationally excited SO lines by \cite{Danilovich2020} and could be due to dynamics in the inner region of R~Dor, where both TiO and vibrationally excited SO originate.

{{The vibrational temperature for TiO in R~Dor was derived by the same procedure adopted for AlO in Sect.~\ref{sec:AlOalma},
except the partition function is from \citet{McKemmish2019}.
Shown  in Fig.~\ref{TiO:RDorvib} is a vibrational temperature diagram for the rotational transitions in the three 
spin components in the ground state and the $v=1$ level, and two spin components in the $v=2$ level.
From the derived $T_{\rm{vib}}$ of 1$800 \pm 170$~K (see diagram in Fig.~\ref{TiO:RDorvib}), we estimate the peak flux of the $J = 11 \to 10$ rotational transitions in the $v = 3$ excited vibrational level ($E_u = 4275$~K) of 19~mJy is within a few mJy of the peak flux in the spectra centered
on the $\Omega = 2$ and $\Omega = 3$ spin components (see Table~\ref{tab:TiOalma} and Fig.~\ref{fig:TiOprofiles}).}}




We conclude that the identifications of the lines of vibrationally excited TiO towards R~Dor are secure and no lines are missing.
The calculated frequencies in the $v=1,~{\rm{and}}~2$ levels reproduce the frequency of all five astronomical features 
to within $\pm 0.3$~MHz of the mean central velocity $\upsilon_{\rm{cent(TiO)}} = 7.8$~km~s$^{-1}$.
The peak flux of the lines in the three spin components in each excited vibrational level are comparable:
$2.4 \pm 0.3$ times lower in the $v=1$ level, $3.3 \pm 0.8$ lower in $v=2$, and $15.1 \pm 0.3$ lower in $v=3$ than the 
corresponding transitions in the ground vibrational state, as expected if TiO is close to the central star and 
$T_{\rm{vib}} = 1800$~K \citep[cf. $T_\mathrm{eff} = 2400$~K,][]{Decin2017}.


\begin{figure}
\centering
\includegraphics[width=0.45\textwidth]{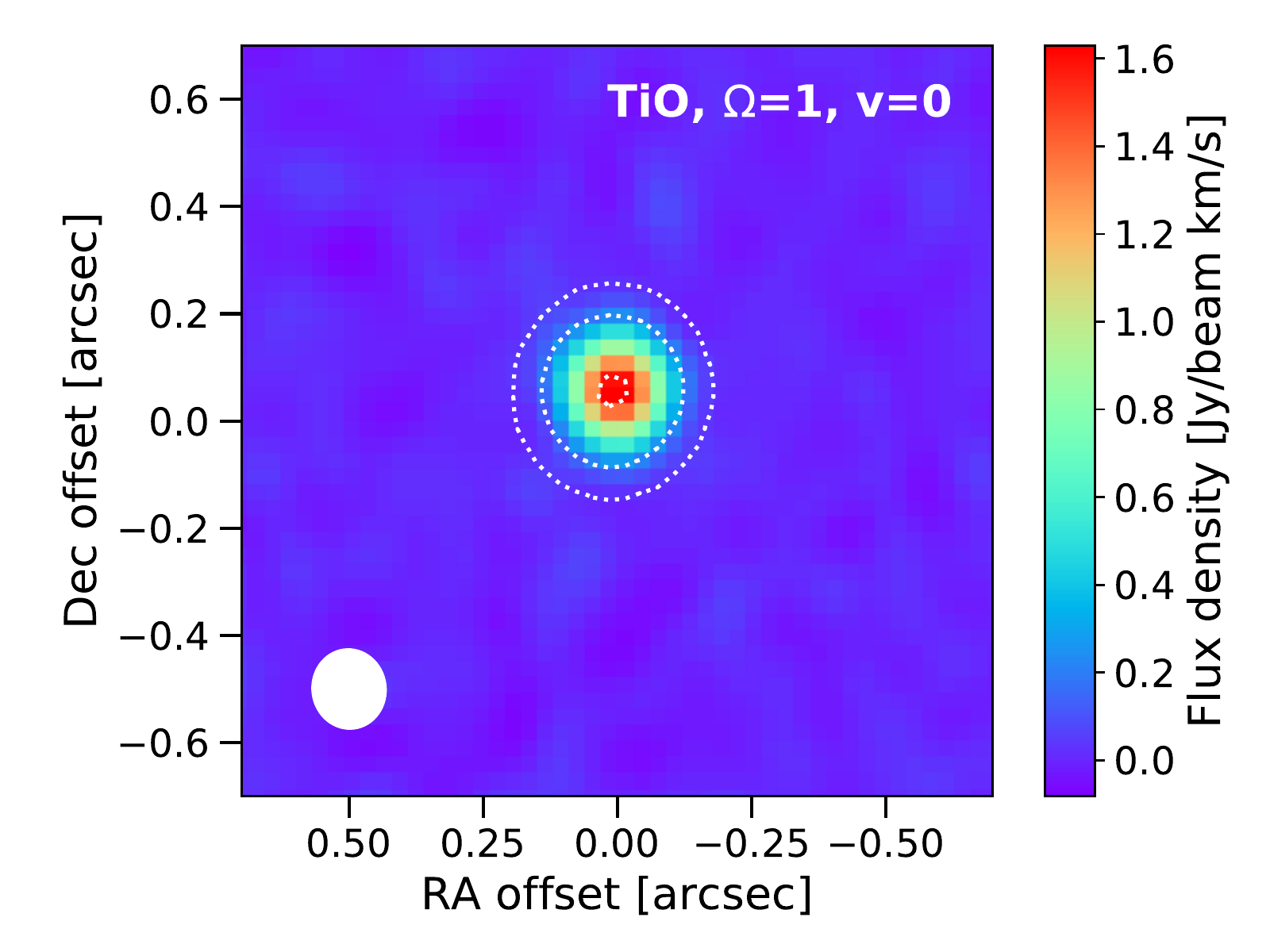}
\caption{\small Zeroth moment map of the lowest fine structure component ($X^3\Delta_1$) of TiO in the ground ($v=0$) 
vibrational state observed in R~Dor with ALMA.
Indicated in white in the lower left corner is the $136 \times 147$~mas beam.
The white dotted contours indicate the continuum flux at levels of 1, 10, and 90\% of peak flux. \label{fig:TiOmap}}
\end{figure}

\begin{figure}
\centering
\includegraphics[width=\columnwidth]{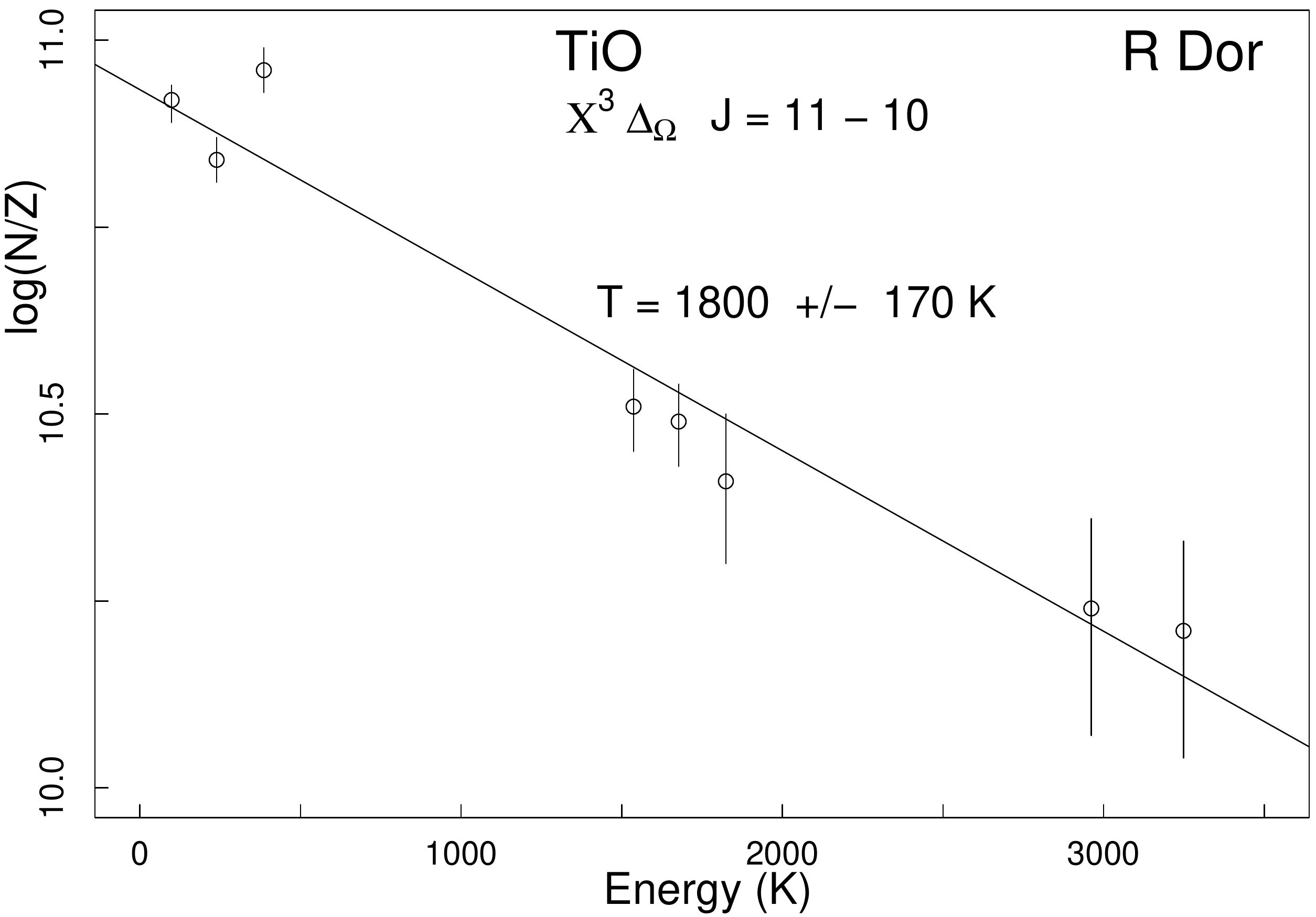}
\caption{\small Vibrational temperature diagram for TiO in R~Dor derived from the $J = 11 - 10$ rotational transition 
in the three spin components in the ground vibrational state and first excited $v=1$ vibrational level.   
Only two rotational transitions were observed in the $v=2$ level, because of the coincidence of the transition in the 
$^3\Delta_2$ component with a line of $^{34}$SO$_2$.}
\label{TiO:RDorvib}
\end{figure}


\begin{figure}
\centering
\includegraphics[width=\columnwidth]{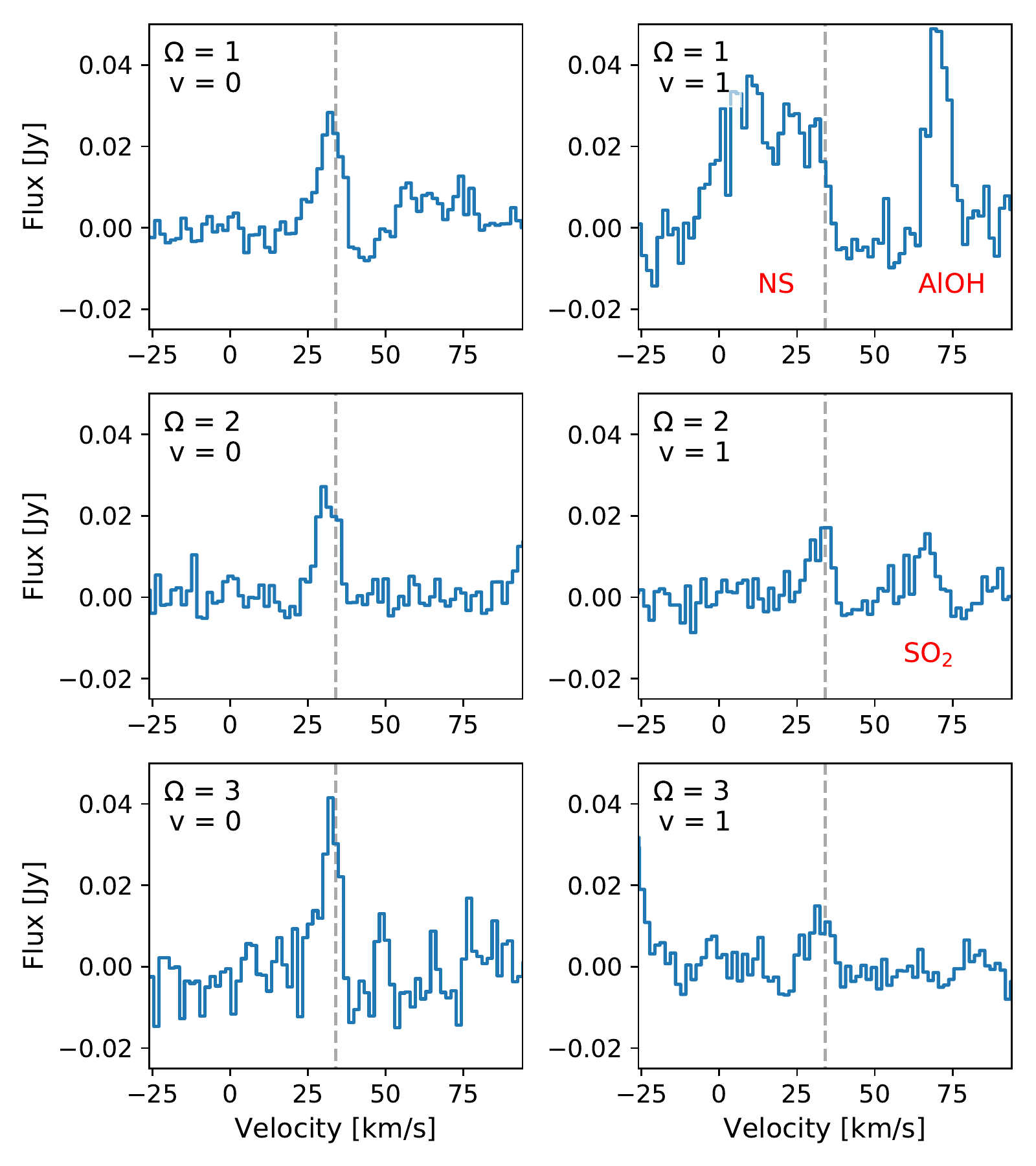}
\caption{\label{fig:TiOiktau} \small  The $J=11\to10$ rotational transition of TiO in the three fine structure components in the ground ($v=0$) and 
first excited vibrational level ($v = 1$) observed towards IK~Tau. 
The spectra were extracted from the ALMA data with an aperture whose radius is 160~mas. 
The grey dashed vertical lines indicate the expected center of the TiO line profiles for the assumed $\upsilon_\mathrm{LSR}=34$~\kms. 
Lines of other molecules are indicated in red.
}
\end{figure}

For IK~Tau the peak flux densities of the lines of TiO in the ground vibrational state are about 10 times lower than for R~Dor 
(see Figures~\ref{fig:TiOiktau} and \ref{fig:TiOprofiles}, and 
Table~\ref{tab:TiOalma}).
Lines in the $v = 1$ level in the $\Omega = 2$ and $\Omega = 3$ spin components were observed with a peak flux 
that is only two times higher than the rms noise, and no lines were detected in the $v = 2$ level of TiO towards IK~Tau. The $\Omega=1$ component of the line in the $v=1$ state is also undetected, due to a strong overlap with the NS line group at 346.22~GHz. The TiO emission from IK~Tau is slightly blue-shifted with respect to the LSR velocity. Considering only the five detected lines shown in Fig.~\ref{fig:TiOiktau}, we find a mean central velocity of $\upsilon_\mathrm{cent(TiO)}=32.4\pm 0.4$~\kms. This is offset from the $\upsilon_\mathrm{LSR} = 34\pm1$~\kms{} {of IK~Tau \citep{Decin2018}} by 1.6~\kms{} towards the blue, and could also be due to as-yet unknown inner wind dynamics.
However owing to the estimated uncertainties in the rest frequencies 
and the low signal-to-noise of the data, confirmation of the apparent offset in velocity will require more accurate laboratory and more sensitive astronomical measurements.


\section{Radiative Transfer Analysis of R~Dor}  
\label{sec:RadTran}

\subsection{Method}

Due to the quality and availability of observations, we only performed radiative transfer analyses for R~Dor.
For this, we used a 1D spherically symmetric non-LTE radiative transfer program based on the accelerated lambda iteration method 
\citep[ALI;][]{Rybicki1991,Maercker2008} to calculate the abundances and line profiles of AlO and TiO.
The calculated line profiles were then compared with the lines of AlO or TiO observed with ALMA as described in
Sects.~\ref{sec:AlOalma}  and \ref{sec:TiOalma}. 

For modelling purposes, we analyzed spectra extracted with a circular aperture with a radius of {150~mas, for AlO, and} 300~mas, {for TiO,} because this included all the observed emission (see Sect. \ref{almadatadesc} for details).
{When ray-tracing our modelled CSE, we used (circular) model beams the same size as the extraction aperture, to allow for direct comparisons. 
{Our model {circumstellar envelopes} (CSEs) were divided into 30 shells, evenly spaced on a  logarithmic scale from {an inner radius $r_0 = 3.3\times 10^{13}$~cm,} out to {an outer radius} $r_\mathrm{max}=4\times10^{14}$~cm.}
The velocity resolution of the model lines is approximately 0.5~\kms, however the final resolution of the AlO lines, taking hyperfine structure into account, is approximately 1~\kms, to match the resolution of the observations.}
The {observed} AlO lines are significantly broadened by the unresolved fine and hyperfine structure, and the TiO lines are split into three 
fine structure components as discussed in Sects.~\ref{sec:PhysAlO}  and \ref{sec:PhysTiO}, respectively. 
In the following two sections (\ref{alohyperfine} and \ref{tiofine}) we describe our treatment of hyperfine structure for AlO and fine structure for TiO, respectively.

\subsubsection{AlO}\label{alohyperfine}

Explicitly including all the AlO hyperfine levels and the connecting transitions in the radiative transfer analysis is computationally prohibitive, even if the included levels are restricted to a relatively small number of rotational levels (e.g., $N\leq20$). 
Following the procedure in \cite{Decin2017}, AlO was first treated without hyperfine structure to obtain the total integrated line strengths for each observed transition. 
The calculated total integrated flux for each transition was then distributed among the individual hyperfine components
according to their quantum mechanical line strengths \cite[see][]{Yamada1990,Goto1994,Launila1994,Launila2009}. 
These were then summed to obtain the theoretical line profile shape. 
We used Gaussians with a HWHM of 3~\kms, close to the innermost velocity as derived from the sound speed\footnote{{The sound speed was found by assuming an ideal gas at a temperature of 1000~K.}}, to represent the hyperfine components. {In Sect.~\ref{app:hyperfine} we discuss the impact of the choice of widths for the hyperfine components. }

\subsubsection{TiO}\label{tiofine}

To analyze TiO in the most efficient way possible, we first ran a radiative transfer model excluding the fine structure --- i.e., TiO was treated as though it were a closed shell molecule. 
To facilitate this, we took the energy levels and Einstein A coefficients from \cite{McKemmish2019} for the lowest fine structure component $\Omega = 1$ (see the left hand panel of Fig.~\ref{TiO_ELDrevised} for an overview of the fine structure in TiO).
After running a radiative transfer model excluding the fine structure, 
we scaled the amplitude of the calculated line profiles to the relative intensities of the transitions in the three fine structure components
tabulated in CDMS, on the assumption of LTE and a gas kinetic temperature of 1800~K,  
and compared our calculated profiles with the observed spectra. 
For $v>0$ we assumed the same intensity scaling for the fine structure components as those in the ground vibrational state 
(see Sect. \ref{tioresults}).

\subsection{Input parameters}

\subsubsection{Molecular parameters}

A proper non-LTE analysis of the observations of the two metal oxides AlO and TiO requires the following molecular data:  
(1) Einstein~$A$ coefficients for {rotational transitions in} the $v=0$ state;
(2) Einstein~$A$ coefficients for {rotational transitions in} the lower excited vibrational levels;
(3) Einstein~$A$ coefficients for {the rovibrational} $\Delta v=1,2, \cdots$ transitions; and
(4) collisional rates for H$_2$--AlO and H$_2$--TiO.
The Einstein $A$ coefficients for AlO and TiO were from \cite{Patrascu2015} and \cite{McKemmish2019}, respectively.
The collision rates for H$_2$--AlO and H$_2$--TiO have not been calculated. Instead the rates for He-SiO \citep{Dayou2006}
and He--NaCl  \citep{Quintana-Lacaci2016} --- appropriately scaled to account for the difference in mass of these systems {\citep[see][for details]{Schoier2005}} ---
were used as surrogates. 

\subsubsection{Stellar and circumstellar parameters}\label{sec:stelparams}

The circumstellar model in our radiative transfer calculations of AlO and TiO is based on the circumstellar models of R~Dor presented in
\cite{Maercker2016}, \cite{Van-de-Sande2018a}, and \citet{Decin2017}. 
The present model  includes more observations of AlO and a larger number of energy levels and transitions than the one in \citet{Decin2017}.  
The TiO model is based on the same circumstellar parameters as AlO, which are given in Table~\ref{stellarparam}.
{This leaves $f_\mathrm{molecule}$, the inner molecular abundance (relative to H$_2$), as the only free parameter in our models.}

\begin{deluxetable}{lccc}
\tablecaption{Stellar and circumstellar parameters for R~Dor.}
\tablewidth{0pt}
\label{stellarparam}
\tablehead{
\multicolumn{1}{l}{Property} & \multicolumn{1}{c}{Units}               &   \multicolumn{1}{c}{}  &        \multicolumn{1}{c}{R Dor}       
}
\startdata
$R_{\star}$                           &    cm                                                &	                                &	$3.3 \times 10^{13}$    \\
$T_\mathrm{eff}$                                    & K 	                                                &	                                &	2400        \\
$D$                                      & pc 	                                                &		                        &	59		\\
$L_\star$                                   & L$_\odot$ 	                                &	                                &	6500	        \\
$\dot{M}$                            &    $10^{-7} M_{\sun}$~yr$^{-1}$ 	&	 	                        &	 $1.6$ 	\\
$\upsilon_{\infty}$               & km~$s^{-1}$                                    &	                                &	5.7		\\
$\upsilon_0$                       & km~$s^{-1}$                                    &	                                &	3		\\
$\upsilon_{\rm{LSR}}$        & km~$s^{-1}$ 	                                &		                        &	7.0	         \\
$\upsilon_{\rm{turb}}$        & km~$s^{-1}$ 	                                &		                        &	1	         \\
$\epsilon$                           &    $\cdots$                                        &                                      &        0.65            \\
$\beta$                               &    $\cdots$                                        &                                       &        1.5            \\
$r_0$                                  & cm                                                    &	                                &	$3.3 \times 10^{13}$      \\
$r_e$                                  & cm                                                    &	                                &	$1 \times 10^{14}$		\\
\enddata
\end{deluxetable}

The gas kinetic temperature was assumed to follow a power law radial profile \citep[][and references therein]{Decin2006}
\begin{equation}\label{Tgas}
T_{\rm{gas}}(r) = T_\mathrm{eff} \left(\frac{R_{\star}}{r} \right) ^{\epsilon} ,
\end{equation}
with $\epsilon=0.65$, $T_\mathrm{eff}=2400$~K, and $R_\star=3.3\times 10^{13}$~cm. This value of $R_\star$ is derived from the implemented values of $T_\mathrm{eff}$ and $L_\star$, and is slightly larger than the value found by \cite{Norris2012} of $2.4\times 10^{13}$~cm (based on our implemented distance of 59 pc) or that used by \cite{Van-de-Sande2018a} of $2.5\times 10^{13}$~cm, who also used a lower luminosity than ours of $4500~L_\sun$, whereas our luminosity value of $6500~L_\sun$ is that found by \cite{Maercker2016}.

The parameterized $\beta$-type accelerating wind is described by
\begin{equation}\label{eq:vel}
\upsilon_{\rm{gas}}(r) = \upsilon_0 + (\upsilon_\infty - \upsilon_0) \left( 1 - \frac{r_0}{r} \right)^\beta ,
\end{equation}
\noindent where the inner wind velocity $\upsilon_0=3$~\kms{} is set to the approximate sound speed, $\upsilon_\infty=5.7$~\kms{} is the terminal 
expansion velocity, $r_0=R_\star$ is the radius at which the wind is launched, and $\beta=1.5$. The values of $\upsilon_\infty$ and $\beta$ were derived from an analysis of 
CO in R~Dor by \citet{Maercker2016}. 
{We also include a constant turbulent velocity of $\upsilon_\mathrm{turb}=1$~\kms.}
The fractional abundance of AlO and TiO were represented by a Gaussian profile 
\begin{equation}
f(r) = f_\mathrm{molecule} \exp \left(-\left(\frac{r}{r_e}\right)^2 \right)~,
\end{equation}
\noindent centered on the central star, where $f_\mathrm{molecule}$ is the central abundance and $r_e$ is the $e$-folding radius. For both molecules, an $e$-folding radius of $1\times10^{14}$~cm was used. For AlO this is the $e$-folding radius found by \cite{Decin2017}. For TiO we don't have any more detailed spatial information, since the emission is unresolved, but we assume a relatively compact distribution that we expect to be approximately similar to that of AlO.


\subsection{AlO results}\label{aloresults}

The AlO model presented in \cite{Decin2017} was specifically fit to the $N=9\to 8$ line in the ground vibrational state. Using the same model for the newer lines, we found that the $v=0$ lines were reasonably well reproduced, but it was not possible to reproduce the $v=2$ line since the \cite{Decin2017} model only included the ground vibrational state.
When choosing how many vibrationally excited levels needed to be included in our model, we considered the effect of infrared pumping on the population of the rotational levels of AlO.
Infrared emission of the central star has a maximum at around $1-2~\mu$m \citep{Van-de-Sande2018a}, which is close to the term energies of the excited vibrational levels 
of AlO between $v=5$ and 10.
Hence, we included
all levels with $N<20$ and $v\leq10$ as tabulated in ExoMol \citep{Patrascu2015,Tennyson2016}. Extending the molecular data in this way resulted in a well-reproduced $v=2$ line, while still giving well-reproduced $v=0$ lines.

Our best fitting model with the expanded molecular data described above has $f_\mathrm{AlO}=1.7\times10^{-7}$ when using collisional rates based on the \cite{Dayou2006} SiO rates (see Sect.~\ref{sec:colrates} for a more detailed discussion of collisional rates). Our modelled lines are plotted on the observed AlO lines in Fig. \ref{fig:AlO}. 
As an aide for future observations, we have plotted in Fig. \ref{rdorAlOpredic} the calculated line profiles for the $N = 9 - 8$ transition in the $v=1$ and 3 excited vibrational levels derived from the same model. This shows the stark difference in width between transitions in different vibrational states.

The peak abundance we obtained is approximately {3}\% of the solar abundance of Al \citep{Asplund2009}. This is larger than that found by \cite{Decin2017} by a factor of {two}. This is most significantly due to the inclusion of the larger molecular dataset, since more levels are available to be populated and infrared pumping is able to contribute. The new observations of lines in the ground vibrational state provided some additional constraints on the model, but the identification of the $v=2$ line was also crucial for checking the completeness and accuracy of our model. {In addition to the molecular data, we also needed to use a smaller inner radius of $R_\star$, compared with the $\sim2R_\star$ value used by \cite{Decin2017}, to properly reproduce the $v=2$ line.}

\begin{figure}
\centering
\includegraphics[width=\columnwidth]{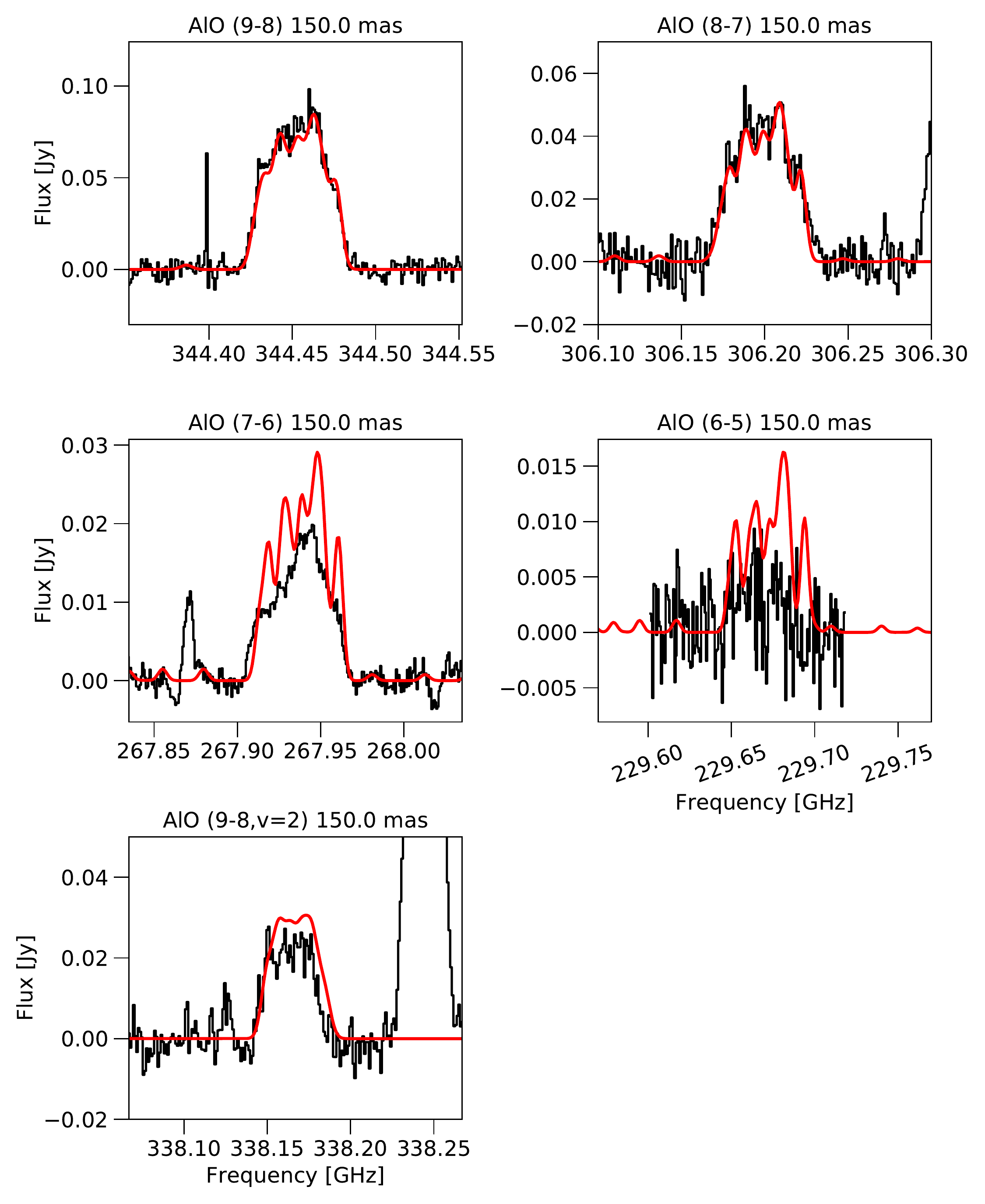}
\caption{Spectra (black histograms) of rotational transitions of AlO observed towards R~Dor, plotted with our synthetic profiles (red curves).
Rotational (and vibrational for $v\neq 0$) quantum numbers and extraction aperture radii are indicated at the top of each box.
The narrow feature at 267.871~GHz is the $15_{3,13} - 15_{2,14}$ transition of $^{34}$SO$_2$, and the intense 
line at 338.245~GHz is the $8 - 7, v=2$ transition of $^{29}$SiO.
}
\label{fig:AlO}
\end{figure}

\begin{figure}
\centering
\includegraphics[width=\columnwidth]{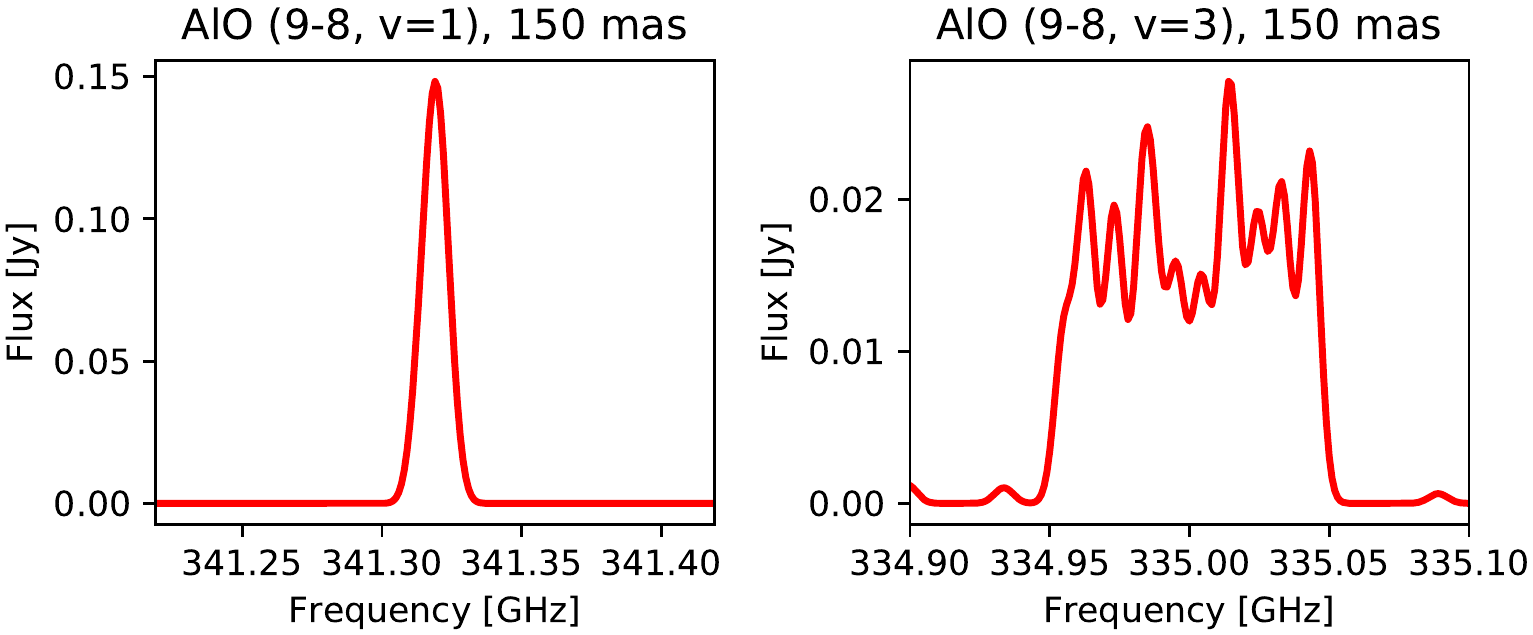}
\caption{Synthetic line profiles of the $N = 9 - 8$ rotational transition of AlO in the $v=1$ and $v=3$ excited vibrational states.
} 
\label{rdorAlOpredic}
\end{figure}

Although the model discussed above reproduces our AlO observations well, we investigated the effects of other factors which might possibly alter our results. In Sect. \ref{sec:colrates} we discuss the effects of different choices of collisional rates. Ultimately, we found that the choice of collisional rates had a small but generally insignificant impact on our model results. {In Sect. \ref{sec:aperture}, we discuss the impact of the choice of spectral extraction aperture on the final model results.} In Sect. \ref{sec:disc} we test the influence of a disk on our model, such as that found by \cite{Homan2018}. We found that including an approximation of a disk had {some} effect on our final results. {{We also tested the impact of an increased infrared radiation field on our results. This is} discussed in detail at the end of Sect. \ref{tioresults} {for TiO. We found that} AlO is not very sensitive to increases in the infrared radiation field.}


\subsection{TiO results}\label{tioresults}

Analogous with our radiative transfer analysis of AlO, we assumed that TiO could be described by a Gaussian distribution 
centered on the star with the same stellar and circumstellar parameters as in AlO, including the same $e$-folding radius 
$r_e =1\times 10^{14}$~cm (see Section~\ref{sec:stelparams} and Table~\ref{stellarparam}).
When adjusting our model to best reproduce the observations, we did not allow the fractional abundance of TiO to exceed the 
solar abundance of Ti \citep[${\rm{Ti/H_2}} =1.8 \times 10^{-7}$;][] {Asplund2009}. 
The initial analysis included rotational levels with $J\leq20$ and vibrational levels with $v\leq 5$, and the velocity profile given by Eq.~\ref{eq:vel}. 
The collisional rates for TiO were derived from the NaCl-He rates for a gas temperature of 700~K  \citep{Quintana-Lacaci2016}, 
scaled for the difference in mass of the TiO-H$_2$ system {\citep{Schoier2005}}.

\begin{figure}
\centering
\includegraphics[width=0.8\columnwidth]{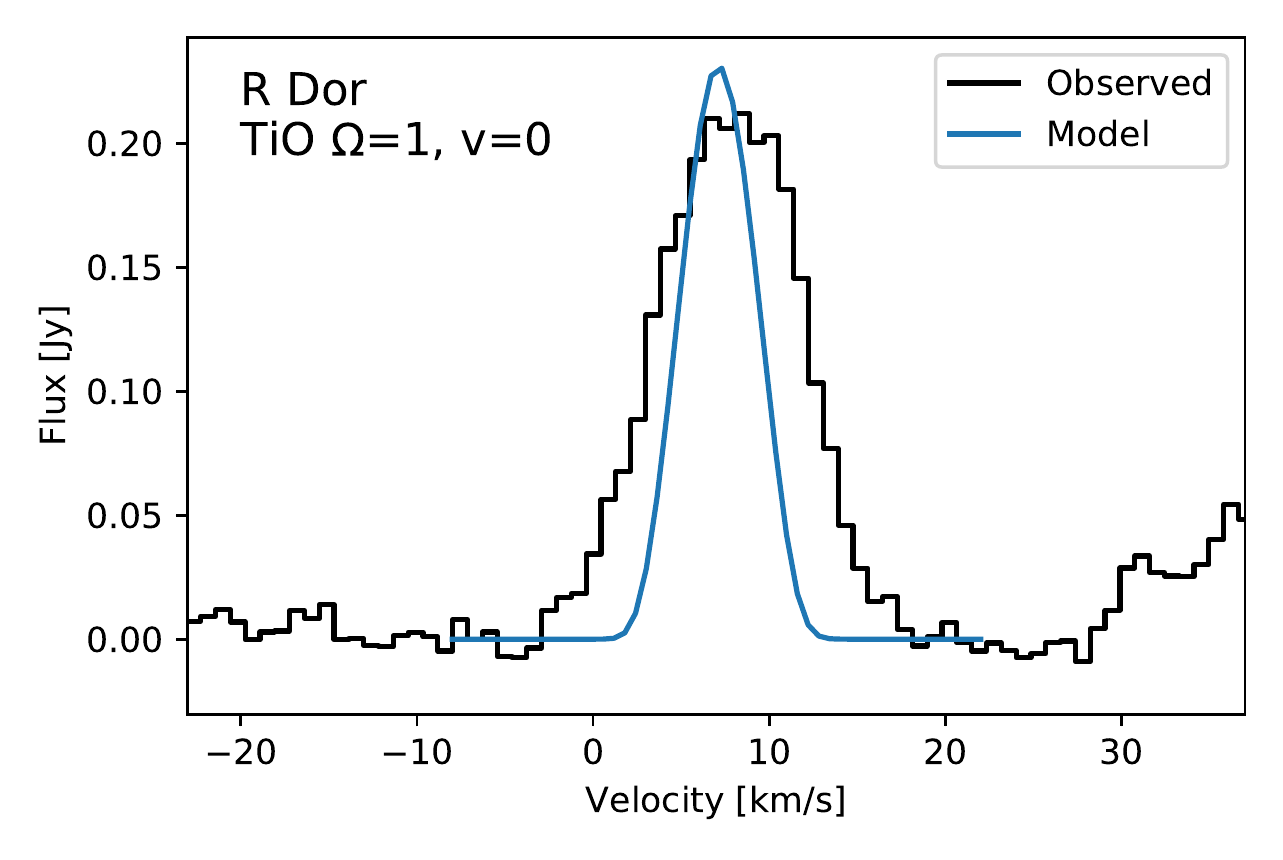}
\caption{\small The $J =11-10$ rotational transition in the $^3\Delta_1$ spin component of TiO in the ground ($v=0$)
vibrational state at 348,159.8~MHz observed towards R~Dor (black histograms).
A theoretical line profile calculated assuming a $\beta$-velocity law 
is shown in blue (see text for details).
}
\label{fig:TiOnarrowline}
\end{figure}

From our first models, it was immediately apparent that the observed TiO lines are much broader than the line widths resulting from the velocity profile given in Eq.~\ref{eq:vel}. A sample line from such a model with $f_\mathrm{TiO} =1.8 \times 10^{-7}$ is shown in Fig. \ref{fig:TiOnarrowline} {and the full set of modelled and observed lines is shown in Fig.~\ref{fig:TiOprofiles_all}}. Although the peak of the $v=0$ model line corresponds well with the observed peak, 
the calculated lines are $\sim50\%$ of the observed $v=1$ peaks and less than $\sim20\%$ of the $v=2$ peaks.
We were able to better reproduce the widths of the observed lines by implementing a constant expansion velocity of 6~\kms{} in place of the $\beta$-velocity law. However, in this case the peaks of the model lines significantly under-predicted the observations with the $v=0$ model lines at $\sim40\%$ of the observed peaks and the $v>0$ model lines at $\lesssim 10\%$. Although it did not have a significant effect on our AlO model, we tested the effect of the ``overdensity" model from \cite{Danilovich2020}, with an increase in H$_2$ number density between $\sim 3R_{\star}$ and $4 R_{\star}$ by a factor of 8. This model, with $f_\mathrm{TiO}=1.0\times 10^{-7}$, produced similar line peaks as the initial velocity law model, including under-predicting the higher-$v$ lines.

For AlO, our modelling could reproduce $v=0$ line when we included a larger number of vibrational states in our model. However, the vibrational levels included in the TiO input molecular data for $v\leq5$ cover a similar extent in energy as those in AlO for $v\leq10$, so we did not expect a significant difference when extending the included levels. Indeed, a test with TiO vibrational levels up to $v\leq10$ confirmed this. Aside from vibrationally excited levels, TiO has many low-lying electronically excited states, for which \cite{McKemmish2019} provides line lists.
The two lowest-lying electronic states are the singlet states $a\,^1\Delta$ and $d\,^1\Sigma^+$ whose term energies of 
$3~\mu$m and $1.8~\mu$m are close to the peak flux of the central star, and are also close to the term energies of the 
$v=3$ and $v=5$ excited vibrational levels of the ground electronic state (X$\,^3\Delta$). We tested models including these two singlet states, and including a larger complement of excited (singlet and triplet) states: $X\,^3\Delta$,  $a^1\Delta$, $d^1\Sigma^+$, $b\,^1\Pi$, $E\,^3\Pi$, $A\,^3\Phi$, $B\,^3\Pi$, $C\,^3\Delta$, and $D\,^3\Sigma^-$. We found that including more electronic levels only caused incremental differences and small improvements in the model, and the model did not adequately reproduce the vibrationally excited lines.\footnote{Our modelling attempts including excited electronic states may have been partly hampered by our inability to include collisional transition rates for those states.} {For clarity, we have included all the discussed models in Table \ref{rdortiomods}.}

\begin{deluxetable*}{clll}[t!]
\tablecaption{Parameters in the radiative transfer calculations of TiO in R~Dor.}
\label{rdortiomods}
\tablehead{ 
\multicolumn{2}{c}{Model designation}
&  \multicolumn{1}{l}{Electronic states}      
&  \multicolumn{1}{l}{Details}   
}
\startdata
I.&$\beta$-velocity law      &  $X\,^3\Delta$    
& Parameters from Table \ref{stellarparam}, $f_\mathrm{TiO}=1.8\times 10^{-7}$, gas velocity from   \\
                              &               &                              &     Eq.~\ref{eq:vel}, rotational levels $J\leq 20$, and vibrational levels $v\leq 5$. \\
II.&$\upsilon_\mathrm{gas} = 6$~km/s            &  $X\,^3\Delta$      &    As in I, 
                                                                                    but with a constant outward gas velocity $\upsilon_\mathrm{gas} =6$~\kms \\
III.&Overdensity                    &  $X\,^3\Delta$      &    As in II, but with the H$_2$ number density 8 times higher for 
                                                                                       \\
                   &                          &                              &     $\sim3$--$4\,R_{\star}$ and $f_\mathrm{TiO}=1.0\times 10^{-7}$.    \\
IV.&Singlet states                 &  $X\,^3\Delta$, $a^1\Delta$, $d^1\Sigma^+$      &   As in III.   \\  
V.&Many excited states  &  $X\,^3\Delta$,  $a^1\Delta$, $d^1\Sigma^+$, $b\,^1\Pi$, $E\,^3\Pi$,    &   As in III.      \\
&&  $A\,^3\Phi$, $B\,^3\Pi$, $C\,^3\Delta$, $D\,^3\Sigma^-$& \\
VI.&Higher IR field   &  $X\,^3\Delta$      &  As in III, but with an additional IR field.   \\  
VII. & Final model   &  $X\,^3\Delta$      &  As in VI, but with $f_\mathrm{TiO}=1.4\times 10^{-7}$.   \\ 
 \enddata
\end{deluxetable*}

All the models discussed until now have not taken dust into account since a significant fraction of the TiO (and AlO) emission comes from within the dust condensation radius found by \cite{Maercker2016}. Including silicate dust with their derived opacity \citep[{optical depth of} 0.05 at 10\micron,][]{Maercker2016} did not have a noticeable effect on the TiO (or AlO) emission. However, there is evidence of a gravitationally bound dust shell \citep{Khouri-thesis2014} or ring \citep{Khouri2016} in the innermost regions around R~Dor, which may overlap with (or be the same object as) the rotating disc seen in molecular lines \citep{Homan2018}. 
Since the dust has been detected with infrared observations, we assume it contributes an additional infrared flux in the
region of interest. To approximate this effect, we included an additional infrared radiation field in our model, based on a black body with a temperature of 1200~K. {Although we have mainly discussed dust as the source of this additional infrared flux, there are other potential sources. In particular, the disc found by \cite{Homan2018} could be caused by a stellar or sub-stellar companion, such as a brown dwarf, which could also be a source of additional infrared flux. Another possible source is stellar fluctuations due to pulsations of the AGB star itself, especially if the stellar parameters used are biased towards fainter phases or epochs. To avoid the problems inherent in choosing just one of these scenarios, we approximate the presence of additional infrared flux using a uniform radiation field which does not vary with distance from the star. {This is done using a source function derived from a black body with a temperature of 1200~K.} This allows us to check the impact of an increased infrared flux, while leaving a more specific examination of the source of such flux to future work, since it is beyond the scope of this present study. With the inclusion of this additional infrared flux,}
we were finally able to reproduce all the TiO lines, including those in higher vibrational states. Our best fit model, which has $f_\mathrm{TiO}=1.4\times10^{-7}$ is shown in Fig. \ref{fig:TiOprofiles}, plotted with the observed spectra. 
Note that this model includes levels with $J\leq 20$, $v\leq 5$ and in the ground electronic state ($X\,^3\Delta$) only. {The other models we tested, as detailed in entries I--VI in Table \ref{rdortiomods} are plotted in Fig. \ref{fig:TiOprofiles_all} in the appendix.}

For completion, we tested a model including the increased infrared radiation field for AlO and found no significant difference between that model and the best model found in Sect. \ref{aloresults}. This is most likely due to the significant difference in Einstein A coefficients for vibrational transitions, which are approximately 30 times larger for TiO than AlO when considering the rotational levels of interest here. This implies that the radiative lifetimes of the excited vibrational levels are shorter for TiO than for AlO, hence a higher infrared flux is needed to maintain sufficient population of vibrationally excited TiO molecules in the higher vibrational levels, to compensate for the shorter lifetimes. {Conversely, the longer radiative lifetimes for AlO mean that an increased infrared flux is not needed to maintain the population of vibrationally excited AlO molecules, and hence does not result in a significantly higher amount of radiative pumping.}

\begin{figure*}
\centering
\includegraphics[width=\textwidth]{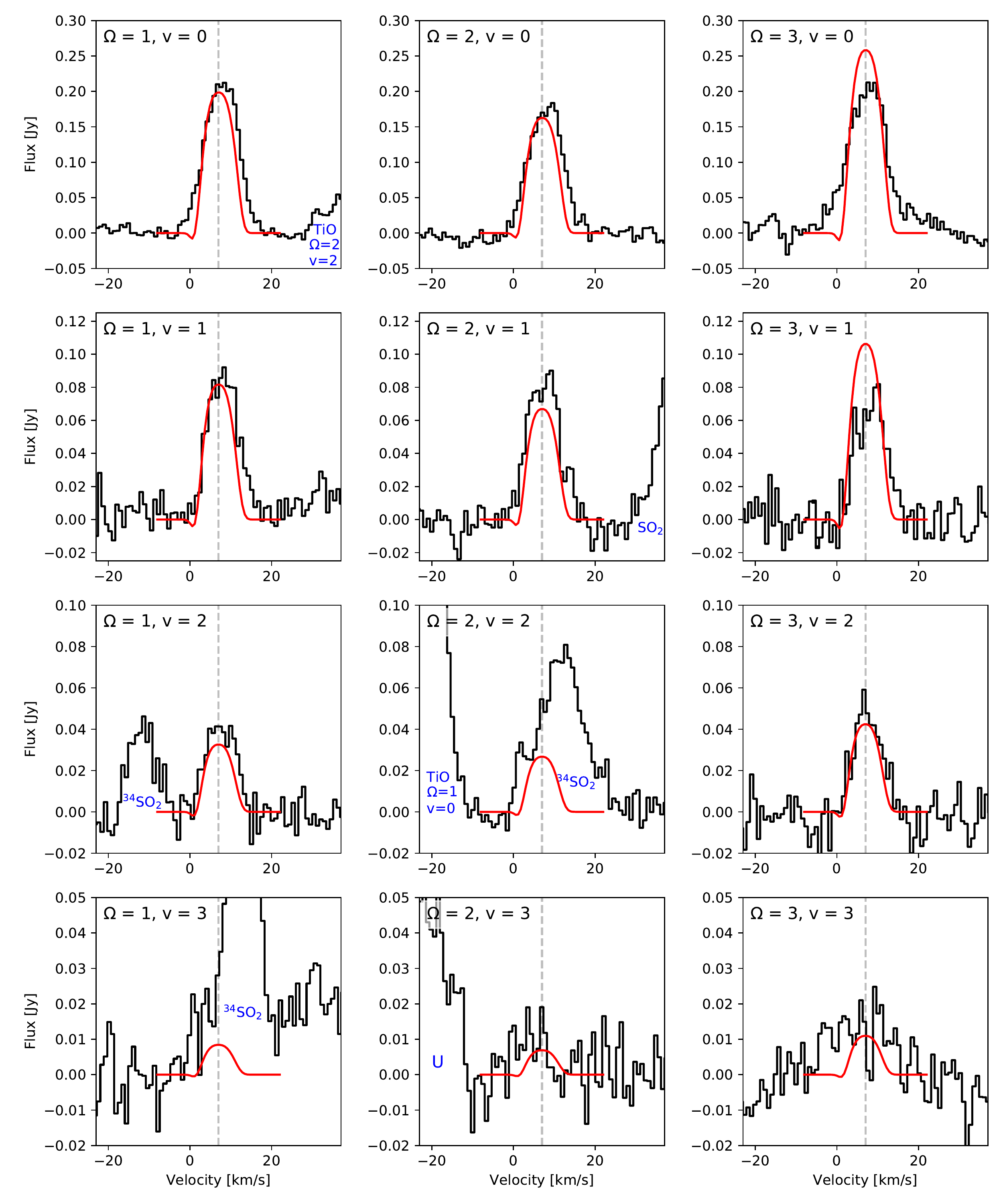}
\caption{\label{fig:TiOprofiles}
The $J =11\to10$ rotational transitions of TiO observed towards R~Dor (black histograms) with the vibrational levels ($v$) and fine structure components ($\Omega$) noted in the top left corner of each plot.
Model profiles are overplotted in red and the vertical dashed grey lines indicate the LSR velocity of 7~\kms. Nearby lines are labelled with blue text.
}
\end{figure*}

\section{Discussion and conclusions}\label{sec:discussion}

\subsection{Our results in context}

Our radiative transfer calculations confirm the importance of infrared radiative pumping on the excitation of AlO and TiO in the inner wind of R~Dor when observed at high angular resolution. For AlO, we found the most important factor was including sufficient vibrationally excited levels to facilitate infrared pumping. However for TiO, we needed to include a sufficiently high infrared flux to reproduce lines in the higher vibrational levels (with $v>0$). Such a flux, however, did not have an effect on our AlO results.

It is apparent from this and prior work on AlO and TiO in R~Dor (see Sects.~\ref{almadatadesc} and \ref{sec:RadTran})
that a substantial fraction of both molecules are located in a region close to the central star  --- 
i.e., $\sim$50\% of gaseous AlO is within the dust condensation radius, and likely a larger fraction of TiO since we only have spatially unresolved observations. A direct determination of the 
size of the compact molecular emission is limited by the angular resolution in the present observations 
\citep[$\sim$140~mas, or 2.5 times the stellar diameter $\theta_{\rm{d}}$;][]{Decin2018}.
As a result, we can only determine the compact molecular emission is from a region within $\sim$70~mas or $\sim 2\,R_{\star}$ 
of the central star \citep[$R_{\star} \simeq  30$~mas][]{Norris2012}. 
However, the radiative transfer analysis of the line profiles has yielded some additional confirmation of the location of AlO and TiO
near and within the dust condensation radius.

\subsubsection{AlO}

Our observations of four successive rotational transitions of AlO in the ground vibrational state and one in the second excited ($v=2$) 
vibrational level are well-reproduced with our radiative transfer model when we included an adequate number of vibrationally excited levels (up to $v\leq 10$, see Sect. \ref{aloresults}). The model reproduces molecular emission which encompasses
the region between the stellar photosphere and the region in which dust starts to condense and extends beyond the dust condensation
radius \citep[see Sect.~\ref{sec:RadTran};][and references therein]{Decin2017,Danilovich2016,Danilovich2020}.

Our revised fractional abundance of AlO in R~Dor ($f_0 = 1.7 \times 10^{-7}$) is about {2} times higher than the 
earlier estimate by \citet[$8.4 \times 10^{-8}$,][]{Decin2017}, but it is still much lower {than the solar abundance of Al}
\citep[$5.6 \times 10^{-6}$, {with respect to H$_2$};][]{Asplund2009}. Hence, this result does not preclude the possibility of AlO being incorporated into dust, even in regions overlapping with our model region.

\subsubsection{TiO}

The line profiles of TiO are wider than expected based on the assumption of the velocity law given in Eq. \ref{eq:vel}, derived from older (less sensitive) single-antenna observations of R~Dor \citep[e.g][who find $\upsilon_\infty=5.7$~\kms]{Maercker2016}. 
Lines wider than the previously determined expansion velocity have been seen for other molecules observed by ALMA. For example, there are faint wings either side of the intense emission lines of CO, SiO, and SO in the ground vibrational state \citep[also observed in the ALMA spectral line survey of R~Dor,][]{Decin2018}. A more detailed analysis of the SO emission by \cite{Danilovich2020} found that the SO lines in the first vibrationally excited state were wider than the HWHM of the $v=0$ lines; that is, rather than just faint wide wings, the $v=1$ lines were overall as wide as the wings of the $v=0$ lines. This was interpreted as emission coming from a rotating disk close to the central star, such as that reported by \citet{Homan2018}, with a rotational velocity greater than the expansion velocity. 
In an LTE analysis of four rotational lines of OH in non-maser emission from high lying rotational lines in R~Dor, 
\citet{Khouri2019} estimated the FWHM of the OH emission region is about $70 \pm 20$~mas with possible evidence for line broadening. This is likely due to the same underlying physical phenomenon. 
The adjustments we made to our TiO model of a higher constant velocity and the inclusion of an overdense region support the rotating disk interpretation. Although the implemented expansion velocity of 6~km~s$^{-1}$ is unphysical in a region so close to the star\footnote{Unphysical under the condition of a smoothly accelerating wind. Other causes of line broadening are also possible, in addition to the rotating disc primarily discussed here, such as infall, expanding shockwaves or standing waves.}, 
it can be thought of here as a partial 1D approximation of the rotating disk.



The presence of photospheric TiO in M-type stars, as seen in bands in the optical and infrared regimes, has long been known and have been used in the spectral classification of these stars \cite[see for example][and references therein]{Gray2009}. Hence, it is not unexpected that we also detect rotational transitions of TiO close to the surface of the star. However, since our observations are not spatially resolved, we are unable to properly constrain the extent of the TiO emission, other than it is within $\sim75$~mas of the star, based on the ALMA beam size of $\sim150$~mas.




Our derived abundance of TiO is $1.4\times 10^{-7}$ relative to H$_2$, which is close to the solar abundance of Ti \citep[$1.8\times 10^{-7}$][]{Asplund2009}. This does not leave much Ti for the possible formation of TiO$_2$ if we assume that TiO$_2$ is present in the same region as TiO. Evidence for the presence of TiO$_2$ is discussed in the following Section \ref{tio2}.
Since the TiO we detect is close to the star and mostly or entirely within the dust condensation radius, our results do not preclude the incorporation of TiO into dust grains. In fact, a role in dust formation or growth would explain the lack of TiO detections further from the star.
It is possible that our apparent high abundance of TiO may rather be attributable to the complex density and velocity structures before the wind acceleration zone in R~Dor, as revealed 
in recent polarimetric observations {in the optical and {near} infrared} at high angular resolution \citep{Khouri2016} and predicted in theoretical dust driven 
models \citep[][and references therein]{Hofner2016}.

\subsection{Evidence for gas-phase TiO$_2$ towards R~Dor}\label{tio2}


\begin{figure}
\centering
\includegraphics[width=0.8\columnwidth]{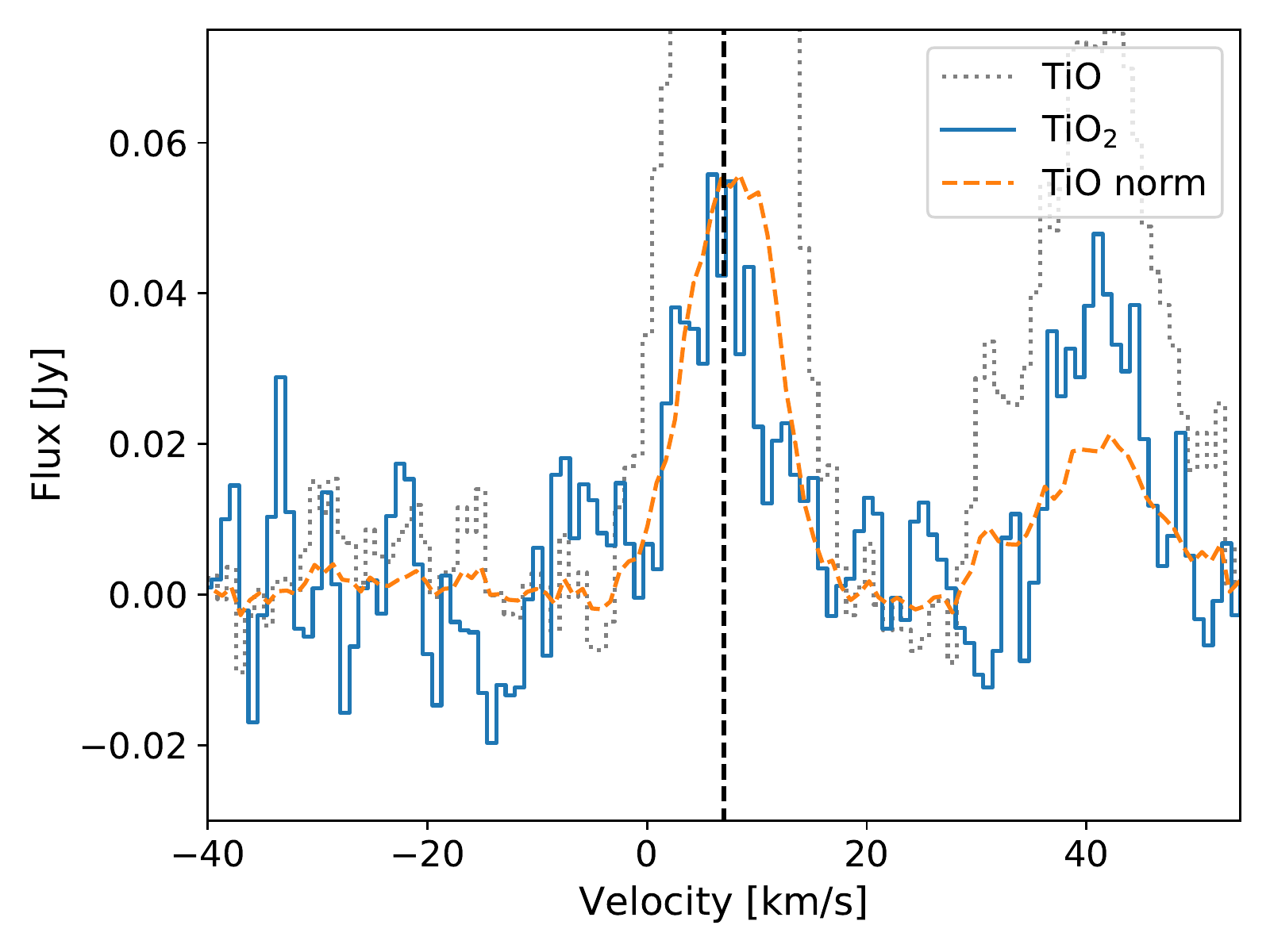}
\caption{The TiO$_2$ ($26_{0,26} \to 25_{1,25}$) line at 350.399~GHz (blue histograms) plotted with the TiO $J=11\to10$, $\Omega = 1$ line in the ground vibrational state (grey dotted line). We also show the same TiO line scaled down to the same peak intensity as the TiO$_2$ line, to allow for a comparison of line profile shapes. The vertical black dashed line represents the systemic LSR velocity of 7~\kms{} for R~Dor.}
\label{fig:TiO2comparison}
\end{figure}

In an earlier survey of R~Dor using the APEX telescope, \cite{De-Beck2018} found some very tentative evidence for TiO$_2$ when stacking 18 lines covered by their frequency range. Using the more sensitive ALMA telescope, \citet{Decin2018} observed some possible TiO$_2$ lines as part of the previously discussed Band~7 spectral line survey of R~Dor (observation ID 2013.1.00166.S). We examined the identified lines in more detail and determined that several lines are uncertain, in large part due to overlaps with other nearby lines. In discussing TiO$_2$ below, we primarily refer to the more certain identifications of lines that we do not believe participate in overlaps, and in particular the following three lines: ($26_{0,26} \to 25_{1,25}$) at 350.399~GHz, ($25_{2,24} \to 24_{1,23}$) at 350.708~GHz, and ($24_{2,22} \to 23_{3,21}$) at 347.788~GHz, listing in descending order of predicted intensity. {We plot one of these lines, ($26_{0,26} \to 25_{1,25}$) at 350.399~GHz with the TiO $v=0$, $\Omega=1$ line in Fig. \ref{fig:TiO2comparison}, showing the TiO line as observed and scaled down to the same peak flux as the TiO$_2$ line.}
The rotational lines of TiO$_2$ are 4--5 times less intense than those of TiO, and the line profile shape {may} be narrower. However, the low intensity {and lower signal to noise ratio} of the TiO$_2$ lines makes it difficult to conduct any conclusive analysis, since even the brightest lines, {such as that shown in Fig. \ref{fig:TiO2comparison}} are only just detected above the noise. The TiO$_2$ emission is also unresolved in our channel maps and zeroth moment maps, supporting the contention that it is located close to the star.
This is feasible since gaseous TiO$_2$ is made by the temperature dependent exothermic reaction   \hfill\break
\begin{equation}
{\rm TiO} + {\rm OH} \rightarrow  {\rm TiO_2} + {\rm H}\, ,
\end{equation}
%
which is very fast at the temperatures within a few stellar radii of oxygen rich AGB stars \citep[see Fig.~2 in][]{Plane2013}.

{An estimate of the column density of TiO$_2$, derived from a single transition in the ground vibrational state 
($26_{0,26} \to 25_{1,25}$) at 350.399~GHz) on the assumption that the excitation temperature is 1800~K, was comparable to the column density of TiO found in our models.
%
However,} this value is very uncertain and does not conclusively imply that there is a similar abundance of TiO$_2$ as TiO (and hence we do not suggest that there is a higher abundance of Ti in R~Dor than in the Sun).

Other related laboratory work concerns the low abundance of TiO$_2$ in the few presolar grains that have been analyzed \citep{Stroud2004,Nittler2008}.
However, owing to the limitation of available samples, the sophisticated laboratory instruments used to analyze the presolar 
grains from low mass AGB stars might not be sensitive enough to detect  TiO$_2$ seed crystals if they were very small 
(L.~Nittler, personal communication).

It has also been noted the abundance of  TiO and TiO$_2$ derived from the rotational spectra in at least one source \citep[Mira A, see][and discussion in Sect \ref{sec:intro}]{Kaminski2017} 
appears 
too low to account for the total mass of the dust in the dust forming region if the dust were composed of TiO$_2$.
Yet there is substantial depletion of Ti in the interstellar medium by about 0.1 or higher 
\citep{Jenkins2009}, suggesting TiO$_2$ might be involved in dust formation.
And as \citet{Boulangier2019} showed, the number of TiO$_2$ clusters formed in their chemical code 
($n({\rm{TiO}}_2)_{1000}/n_H \sim 10^{-11}$) should be sufficient to drive a wind and still be left with a large abundance of 
gaseous TiO$_2$/TiO/Ti.

Unfortunately, rotational lines of vibrationally excited TiO$_2$ have not been observed in the laboratory, nor in oxygen-rich AGB or RSG stars \citep[for a more comprehensive look at TiO$_2$ in VY CMa, see][]{Kaminski2013b}.
This leaves us with few comparisons and little room to draw any definite conclusions regarding TiO$_2$.

\subsection{Dust precursors and dust formation}

{Optical and i}nfrared observations of R~Dor and W~Hya, another oxygen-rich AGB star with a low mass-loss rate,  
established that there are transparent dust grains within $\lesssim 2 R_{\star}$ of the central star 
{\citep{Norris2012, Khouri2016,Ohnaka2016}}.
The composition of the dust grains has not been conclusively confirmed, but the present evidence points to alumina (Al$_2$O$_3$) 
as the probable carrier of spectral features in the mid IR, as observed at low angular resolution with ISO~SWS
\citep[see][and references therein]{Khouri-thesis2014,Decin2017}. 
More recently, these and other oxygen-rich AGB stars have been observed at much higher angular resolution using mid {and near} IR interferometry 
\citep{Karovicova2013,Ohnaka2016}, polarimetric interferometry \citep{Norris2012}, {and polarimetric imaging} \citep{Khouri2016,Ohnaka2016}.

There has been a long standing debate about the formation mechanism(s) of solid Al$_2$O$_3$ grains from gas phase species,
and whether it entails a  homogenous or a heterogenous process. 
\citet{Gail1998,Gail1999} concluded --- on the assumption of chemical equilibrium in the gas phase, approximate nucleation theories, 
and extrapolation of bulk properties to gas phase clusters --- that a heteromolecular process in which small aluminum 
bearing oxides and hydroxides (e.g., AlOH, Al$_2$O, and Al$_2$O$_2$) condense on a grain surface and subsequently react to 
produce the Al$_2$O$_3$ grains appears most probable\footnote{Summarized in Appendix \ref{ap:structures} are the theoretical structures, 
relative energies, and permanent electric dipole moments of the isomers of Al$_2$O, Al$_2$O, and Al$_2$O$_2$ whose rotational spectra have not yet been measured at high resolution in the laboratory or identified by radio astronomers.}.
This is because the abundance of gaseous Al$_2$O$_3$ is predicted to be far too low to form solid 
Al$_2$O$_3$ by homogenous nucleation of gaseous Al$_2$O$_3$ and subsequent condensation on TiO$_2$ nucleation seeds \citep{Jeong2003}.

There have also been some important advances in supporting laboratory measurements in the past 20 years. 
These include measurements of the IR-REMPI spectra of large gas phase (Al$_2$O$_3$)$_n$ and (TiO$_2$)$_n$ clusters 
which provide accurate wavelengths of the vibrational bands in the IR, and establish Al$_2$O$_3$ and TiO$_2$ clusters 
can form in the gas phase under the specific conditions in the laboratory  \citep{Demyk2004},
but it is uncertain whether the same conditions apply within a couple of stellar radii of the low mass oxygen rich AGB stars.

In a recent study, \citet{Boulangier2019} examined four precursor candidates of the dust with their self-consistent wind model 
of oxygen rich AGB stars: TiO$_2$, MgO, SiO, and Al$_2$O$_3$. 
On the basis of theoretical quantum chemical calculations, they found there is a sharp threshold in temperature of 1000--1200~K 
when (TiO$_2$)$_{10}$ clusters form in their model AGB star; 
the abundance of the (TiO$_2$)$_n$ clusters is orders of magnitude lower at temperatures greater than 1200~K; 
Al$_2$O$_3$ clusters are predicted at temperatures as high as 1800--2400~K; and 
nearly all the available Al$_2$O$_3$ monomers are tied up in the clusters at slightly lower temperatures of $1600 - 2200$~K.
\citet{Boulangier2019} ruled out TiO$_2$ clusters as nucleation seeds,
because the observations establish the dust exists close to the central star where the temperatures are as high 
as $1500 - 2000$~K.
However the uncertainties in the quantum calculations of the temperature ranges are difficult to quantify.
Also as they note, their ``comprehensive'' chemical kinetic code, which starts with atomic constituents, has several significant shortcomings: 
it includes a relatively modest chemical network ($\sim330$ reactions)
and,
although TiO$_2$ and its clusters are produced, their kinetic code does not form the Al$_2$O$_3$ monomer or its clusters; 
AlO$_2$ and Al$_2$O$_2$ --- the suspected precursors of Al$_2$O$_3$ ---  are not produced; and
Mg remains in the atomic form and does not form MgO.

In view of the current understanding of dust formation, further astronomical observations, and accompanying radiative transfer analysis of the suspected gas phase 
precursors in the ground and excited vibrational levels at high angular resolution and sensitivity are needed in conjunction with:
polarimetric {observations in the optical and {near} IR,}
theoretical calculations, 
supporting laboratory measured reaction rates and product distributions, and laboratory measurements 
of the rotational spectra of key species in the kinetic models.
At present the relevant astronomical observations are available for just a couple of specific sources, and there is insufficient 
information to determine whether what we see are general trends in these types of sources.
For example, it is unclear whether gravitationally bound dust shells are prevalent in oxygen rich AGB stars with low mass loss rates 
as suspected \citep{Hofner2018}, or
whether the Al$_2$O$_3$ dust (and by inference gaseous AlO) and gravitationally bound dust shells are intimately connected 
in these objects, 
or whether TiO$_2$ are the seeds of the alumina dust.

\acknowledgments

We are very grateful for helpful discussions on the molecular physics of TiO by Peter Bernath, 
elemental depletion by Tom Millar, and presolar grains by Larry Nittler. We also thank the anonymous referees for constructive comments.
TD acknowledges support from the Research Foundation Flanders (FWO) through grant 12N9920N.
CAG and KLKL acknowledges partial support from NSF grant AST-1615847.
LD acknowledges support from the ERC consolidator grant 646758 AEROSOL and the FWO Research Project grant G024112N.
TK acknowledges funding from grant no 2018/30/E/ST9/00398 from the Polish National Science Center.
This paper makes use of the following ALMA data: ADS/JAO.ALMA2013.0.00166.S. ALMA is a partnership of ESO (representing its member states), NSF (USA) and NINS (Japan), together with NRC (Canada) and NSC and ASIAA (Taiwan), in cooperation with the Republic of Chile. The Joint ALMA Observatory is operated by ESO, AUI/NRAO and NAOJ.
The SMA is a joint project between the SAO and ASIAA and is funded by the Smithsonian Institution and the Academia Sinica.


\bibliography{AlO_TiO_2020}{}
\bibliographystyle{aasjournal}

\appendix

%
%
%
%

%


%

\section{Further detailed investigation of A\lowercase{l}O modelling}

\subsection{Choice of collisional rates for AlO}\label{sec:colrates}

Since the cross sections for collisions between AlO and H$_2$ have not been measured or calculated, we tested the dependence of our radiative transfer model on the choice of collisional rates.
We implemented collisional rates calculated for two other molecular systems: 
(1) SiO colliding with He which only includes levels in the ground vibrational state \citep{Dayou2006}; and
(2) NaCl with He which includes rotational levels in the ground and vibrationally excited states, and ro-vibrational collisional transitions \citep{Quintana-Lacaci2016}.
In both systems, the collisional rates were scaled for the difference in mass with that of AlO--H$_2$ \citep{Schoier2005}. 

To test the influence of the collisional excitation rates on the solutions to the statistical equations in the radiative transfer analysis,
we also ran models with (3) the collision rates adapted from the NaCl rates multiplied by 100, and (4) all the collision rates set to zero.
To compare the dependence of the models on the different collisional rates, we first determined the best fitting model with
the collisional rates adapted from the SiO rates ($f_\mathrm{AlO}=1.7\times10^{-7}$), and then used the same AlO abundance 
profile to run models with the other sets of collisional rates. 
Plotted in Fig.~\ref{fig:AlOallcols} are all the calculated line profiles superposed on the observed spectra.

The difference between the calculated lines profiles and the observed lines for the different sets of collisional rates was negligible in the 
ground vibrational state and was small for the line in the excited $v=2$ level. 
The most significant departure was for the model with the collisions adapted from NaCl multiplied by 100, which resulted in only a minor 
difference for the $v=0$ lines, but produced a profile for the $v=2$ line that was 30\% less intense than the corresponding line calculated 
for NaCl collisions cross sections that were not multiplied by 100.
The model which neglected collisional excitation predicted slightly brighter lines for all observed transitions than the models for which 
collisions were considered.

\begin{figure}[h]
\centering
\includegraphics[width=\columnwidth]{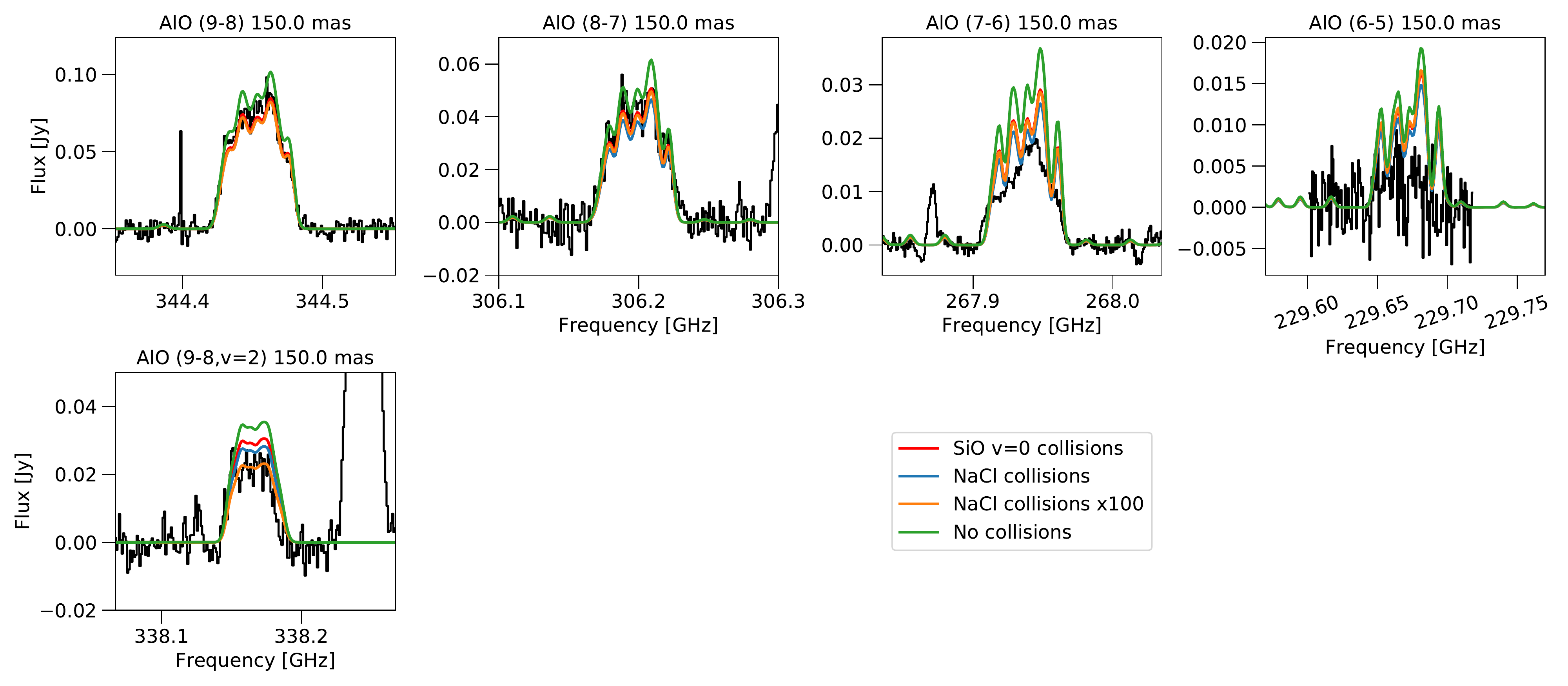}
\caption{Spectra (black histograms) of four successive rotational transitions of AlO in the ground vibrational state and one transition
in the second excited ($v=2$) level observed towards R~Dor; 
and synthetic profiles (coloured curves).
Rotational quantum numbers and the extraction aperture radii are indicated at the top of each box.
}
\label{fig:AlOallcols}
\end{figure}

\subsection{Impact of the south-eastern extension in AlO emission}\label{sec:aperture}
{When selecting spectra to use to constrain our 1D radiative transfer models of R~Dor, we selected AlO spectra extracted for a 150~mas radius aperture. This is in contrast with the 300~mas aperture spectra selected for TiO, which were chosen to capture the maximum amount of TiO flux, without being dominated by noise. In the case of AlO, a smaller aperture was selected to avoid including the SE extension seen in the zeroth moment maps of Fig~\ref{fig:AlOmaps}. The extension breaks spherical symmetry and hence cannot be reproduced by a 1D model.}

{To emphasise the impact the SE extension and our choice of aperture, we plot in Fig. \ref{fig:AlO300mas} the same model as shown in Fig. \ref{fig:AlO} but for a spectral extraction aperture of 300~mas. The differences between Figures \ref{fig:AlO} and \ref{fig:AlO300mas} clearly show the impact of the SE extension: the ground state lines of ($9\to8$) and ($8\to7$) are slightly under-predicted when additional emission from the SE extension is included in the observed spectra; and the ($7\to 6$) and ($6\to5$) lines are less over-predicted for the same reason. In addition, the 150~mas extraction of the ($6\to5$) is strongly affected by resolved-out flux, which is a less prominent problem for the 300~mas aperture spectrum, since the latter is more strongly dominated by noise.
}

\begin{figure}[h]
\centering
\includegraphics[width=\columnwidth]{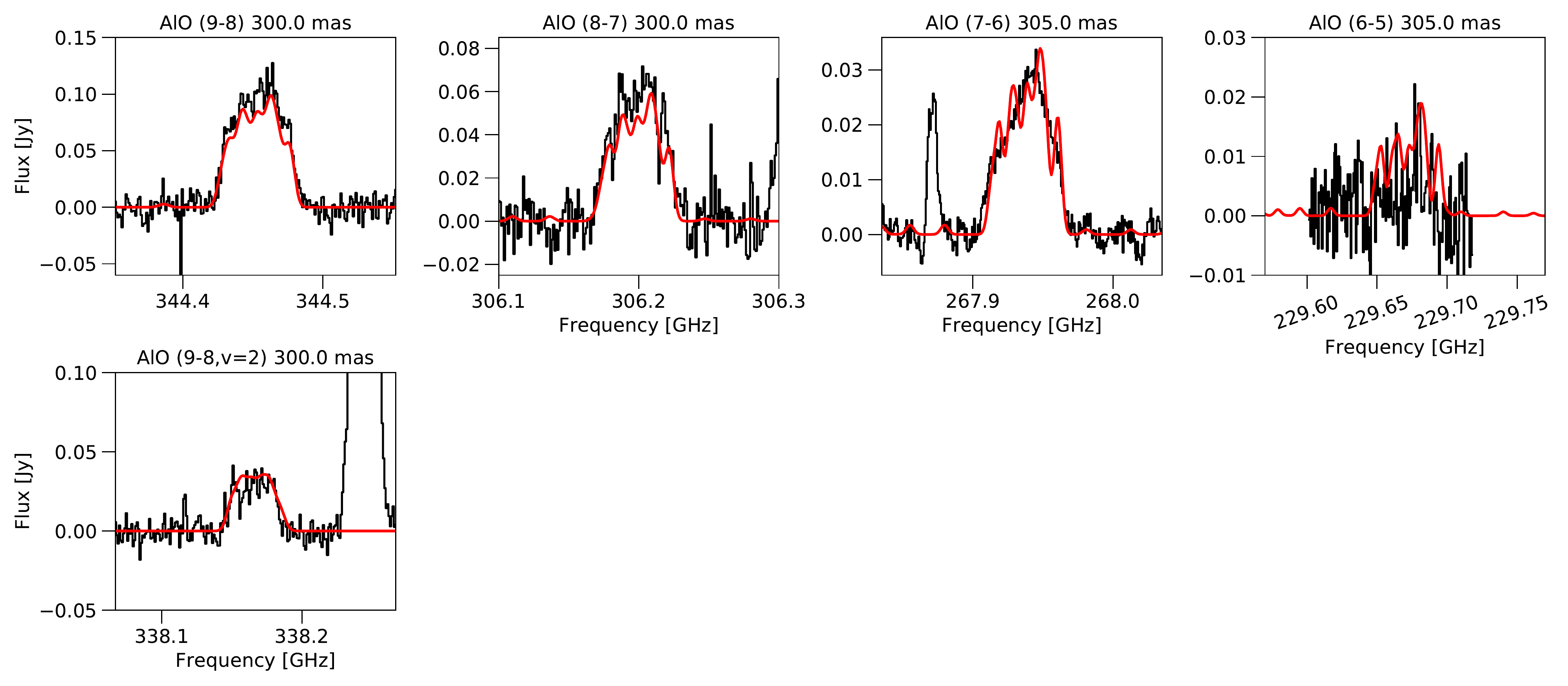}
\caption{Spectra (black histograms) of four successive rotational transitions of AlO in the ground vibrational state and one transition
in the second excited ($v=2$) level observed towards R~Dor; 
and synthetic profiles (red curves) calculated with our best radiative transfer model. Data and models are shown for a spectral extraction aperture radius of $\sim300$~mas.
}
\label{fig:AlO300mas}
\end{figure}

\subsection{The effects of a disk}\label{sec:disc}

In addition to investigating the impact of the particular set of collisional rates adopted, we checked the effect of a possible disk in the inner regions of R~Dor \citep[see][]{Homan2018} on our radiative transfer 
analysis of AlO.
Because our model is 1-dimensional and cannot fully reproduce a disk, we referred to the spherically symmetric approximations implemented by \citet{Danilovich2020} in their analysis of SO around the same star. 
Specifically, we tested the effect of their ``overdensity" model in which the H$_2$ number density between 
$\sim 3R_{\star}$ and $4R_{\star}$ was increased by a factor of 8. 
{Due to }
the overlap between this overdensity region and the region where AlO is primarily observed, {the new model does result in brighter AlO line profiles. Using this overdensity model, we adjusted our AlO abundance until we were able to once again reproduce the observed lines. From this we found a new best fit model with $f_\mathrm{AlO} = 7\times 10^{-8}$ relative to H$_2$. This is more than a factor of two lower than our original model that does not include an overdense region.}

\subsection{Widths of hyperfine components}\label{app:hyperfine}

{In addition to the above, we tested the effect of the assumed 
widths of the theoretical Gaussian profiles representing the unresolved hyperfine components of the very broad AlO lines. 
In general, smaller assumed widths resulted in calculated line profiles in which the underlying hyperfine structure is more 
noticeable, while larger widths resulted in smoother line profile shapes. 
Although we present our final model with HWHM widths set to 3~\kms, we tested different HWHM widths for all the lines. We found the theoretical line profiles which yielded the best fits were 3~\kms{} 
for the $N=9\to8$ and $8\to7$ lines, and 4.5~\kms{} for the $7\to6$ line. The $6\to5$ line is omitted from this analysis due to lower data quality, but generally appears to correspond better with a larger HWHM. 
The tightest observational constraints were found for the $N=9\to8$ and $7\to6$ transitions in the ground vibrational state, 
where wider Gaussian components produced a smoother line profile than was observed for the $9\to8$ line, 
and narrower components produced a jagged line profile that did not reproduce the $7\to6$ line well. As shown in Fig.~\ref{fig:hyperfinex}, the profile constructed from the 5.7~\kms{} hyperfine components does not reproduce the hyperfine structure observed with ALMA for the $9\to8$ line, whereas the the profile constructed from 3~\kms{} hyperfine components (also shown in Fig. \ref{fig:AlO} for all lines) is more jagged than the observed $7\to6$ profile. }

{The good agreement between most of the observed and synthetic line profiles for the 3~\kms{} hyperfine components is part of the reason this value was chosen for the final model. However, we cannot easily explain why the $7\to6$ line has smoother or less distinct hyperfine components. One possible explanation is that, as a lower-energy line, it is formed further out in the envelope. However, the zeroth moment map of the emission (see Fig. \ref{fig:AlOmaps}) and the emission extents predicted by our (spherically symmetric) radiative transfer model do not support this hypothesis. (Note that the apparent larger size of the emission shown in Fig. \ref{fig:AlOmaps} is mainly due to differences in sensitivity between the $7\to6$ and $8\to7$ observations.) Our radiative transfer model cannot account for non-spherically symmetric features, so it is possible the $7\to6$ line is excited more in (the faster rotating portion of) the disc than the other transitions. This is not something we can check with our current level of observations, however.}


\begin{figure}[h]
\centering
\includegraphics[width=0.6\textwidth]{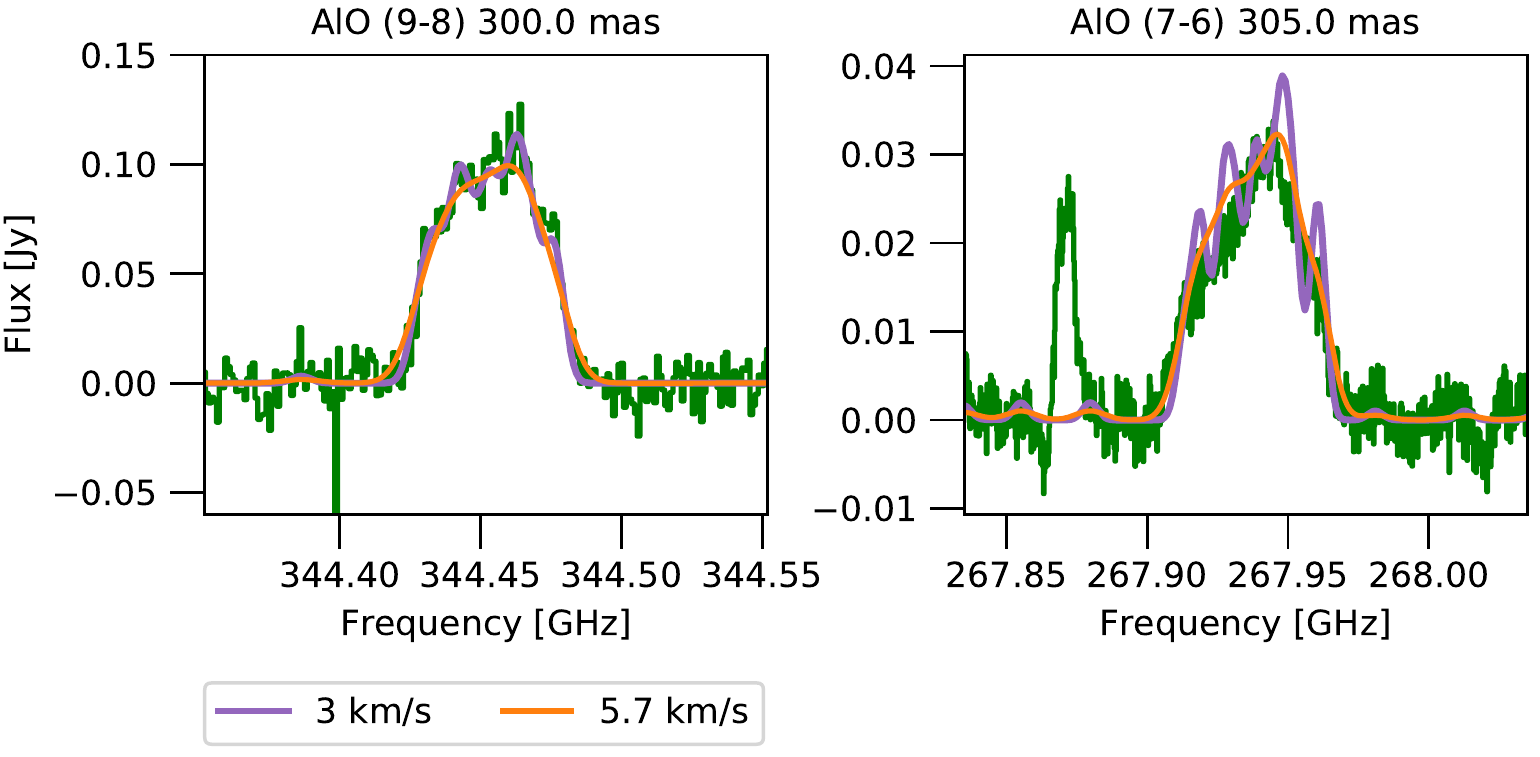}
\caption{
Synthetic line profiles for two transitions of AlO towards R~Dor calculated using two different HWHM (3~km~s$^{-1}$ and 5.7~km~s$^{-1}$) for the Gaussians representing the hyperfine components. These curves are plotted over the observed line profiles, which are shown as green histograms.}
\label{fig:hyperfinex}
\end{figure}

\section{Further details of tested T\lowercase{i}O models}

{In Fig. \ref{fig:TiOprofiles_all} we plot the TiO models discussed in Sect. \ref{tioresults} and tabulated in entries I--VI of Table \ref{rdortiomods}. Our final TiO model, entry VII in Table \ref{rdortiomods}, is not included here and can instead be seen in Fig. \ref{fig:TiOprofiles}.}

\begin{figure*}
\centering
\includegraphics[width=\textwidth]{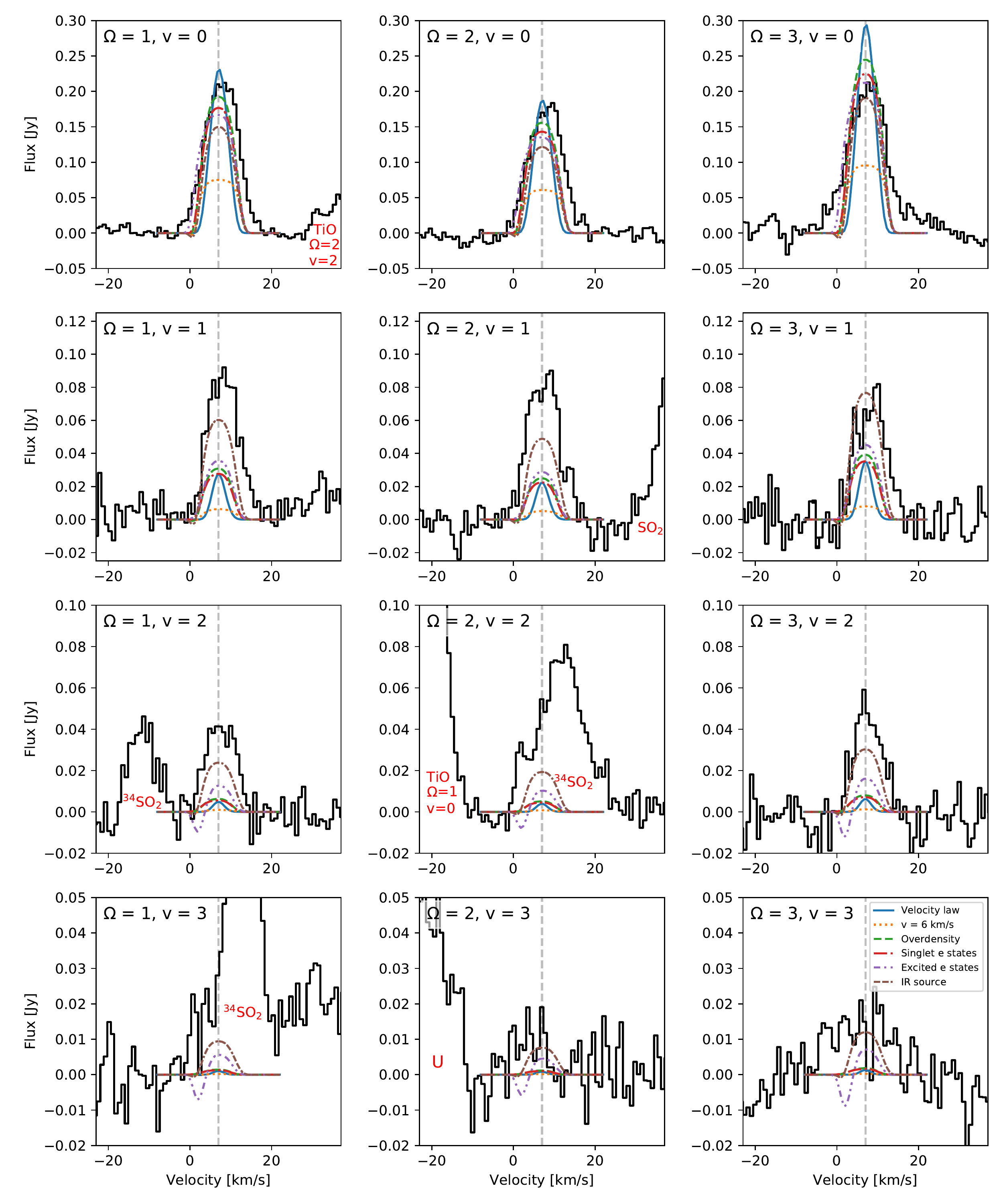}
\caption{\label{fig:TiOprofiles_all}
The $J =11-10$ rotational transitions of TiO observed towards R~Dor (black histograms) with the vibrational levels ($v$) and fine structure components ($\Omega$) noted in the top left corner of each plot.
Model profiles are overplotted in red and the vertical dashed grey lines indicate the LSR velocity of 7~\kms. Nearby lines are labelled with red text.
}
\end{figure*}

\section{Theoretical structures of small aluminum oxides}\label{ap:structures}

There has been some confusion in the astronomical literature regarding the structures of the most stable forms of the
small aluminum oxides, owing to earlier calculations in the chemical literature at low levels of theory.
Presented here are couple cluster theory calculations of three aluminum oxides frequently cited as possible precursors
of Al$_2$O$_3$: Al$_2$O (Fig. \ref{Fig:Al2O}), AlO$_2$ (Fig. \ref{Fig:AlO2}), and Al$_2$O$_2$ (Fig. \ref{Fig:Al2O2}).


\subsection{Al$_\mathit{2}$O}

There are three possible arrangements of the nuclei in Al$_2$O (a $C_{2v}$ form like H$_2$O, a $D_{\infty h}$ form like CO$_2$, and 
a $C_{\infty v}$ form Al--Al--O), and two possible electron spin multiplicities (singlet and triplet). 
Summarized in Figure~\ref{Fig:Al2O} are the calculated structures, permanent electric dipole moments, and relative energies 
at the ae-CCSD(T)/cc-pWCVTZ level of theory, where the zero-point energy was estimated with a smaller basis (cc-pWCVDZ). 

The Al$_2$O isomers with triplet multiplicity are at least 33,000~K or higher in energy than the singlet structures.
Singlet Al--O--Al is the most stable isomer and is probably the most abundant isomer in the inner wind of oxygen rich AGB stars,
but it is nonpolar and the rotational spectrum is unobservable with radio telescopes. 
Two of the  remaining three isomers possess large dipole moments and have structures with $C_{\infty v}$ symmetry ---
one is a singlet and the other is a triplet. 
The triplet $C_{2v}$ form is calculated to be the lowest energy polar form of Al$_2$O, although it might prove difficult 
to detect because it possesses a small dipole moment (0.8~D), and it is nearly degenerate (to within ${\sim}2500$\,K) with the 
$D_{\infty h}$ structure.
Depending on how well the relative energies are determined, the $D_{\infty h}$ structure might undergo thermal 
interconversion between the polar and nonpolar forms in the hot inner wind close to the central star, and therefore warrants 
a more sophisticated quantum chemical treatment.


\begin{figure}[h]
\centering
\includegraphics[scale=0.75]{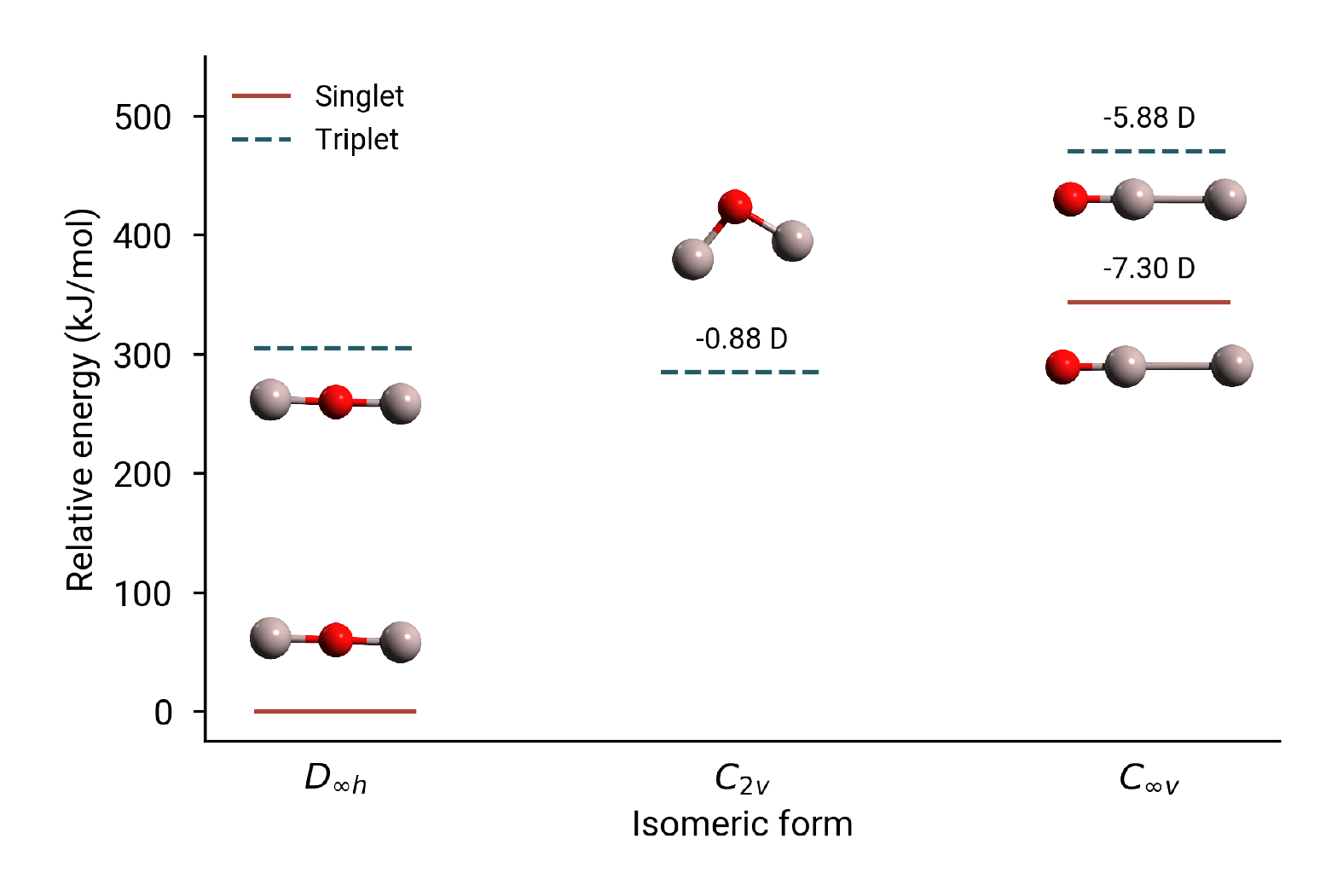}
\caption{\label{Fig:Al2O} 
Relative energies and dipole moments of the isomers of Al$_2$O.
}
\end{figure}


\subsection{AlO$_\mathit{2}$}

To assess whether the rotational spectrum of AlO$_2$ (all with doublet electron spin multiplicity) might be detectable with radio telescopes, we calculated the electronic and zero-point energy, and the dipole moments at the ae-CCSD(T)/cc-pWCVTZ level of 
theory (Figure~\ref{Fig:AlO2}).
The most stable isomer is the nonpolar $D_{\infty h}$ structure, followed by the isomer with $C_{2v}$ symmetry, 
and a linear $C_{\infty v}$ isomer. 
The dipole moments of the polar isomers are modest, perhaps indicating their detection in the inner wind of oxygen rich AGB stars
might be difficult. 
The calculated quadrupole coupling constant ($eQq$) of $-20.9$~MHz for the $C_{2v}$ isomer and $-26.6$~MHz for the $C_{\infty v}$
isomer are similar to that of AlO  \citep[$eQq = -26.1 \pm 0.21$~MHz;][]{Goto1994}, implying the Al---O bonding character is very 
similar in AlO and AlO$_2$.


\begin{figure}[h]
\begin{center}
   \includegraphics[scale=0.75]{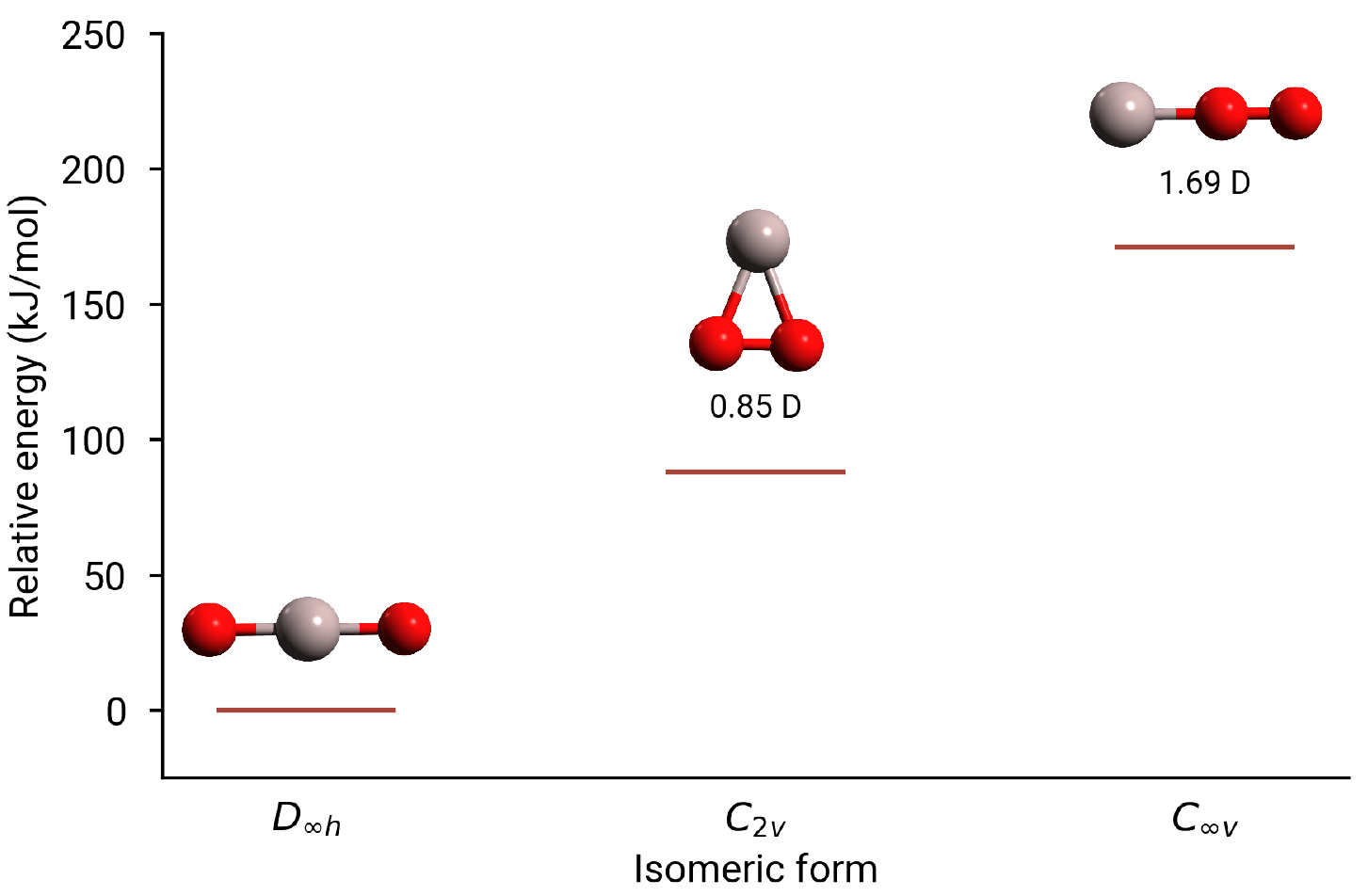}
\caption{\label{Fig:AlO2} Relative energies and dipole moments of three isomers of AlO$_2$ with electron 
spin multiplicity of 2.
}
\end{center}
\end{figure}





\subsection{AlOAlO}

Linear AlOAlO (see Fig. \ref{Fig:Al2O2}) is an attractive candidate for astronomical detection in oxygen rich stars.
It has been observed at IR wavelengths trapped in an inert matrix \citep{Andrews1992},
and the permanent electric dipole moment is large (5.16 D; S. Thorwirth, personal
communication).  Owing to its rigid linear geometry, the rotational spectrum of AlOAlO
consists of a single series of harmonically related rotational transitions spaced every
3.4 GHz apart.

\begin{figure}
\begin{center}
%
      \includegraphics[scale=1.0]{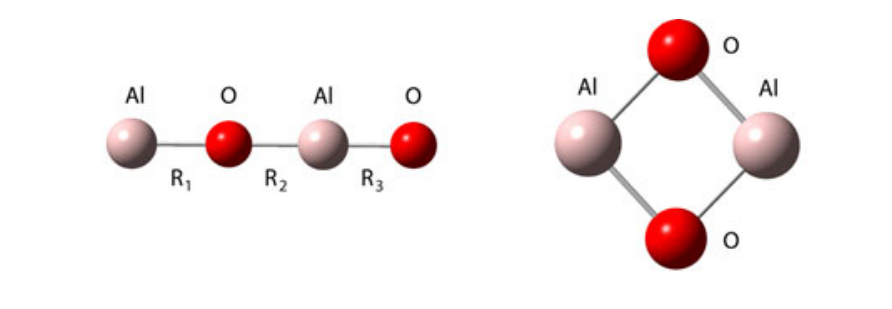}
\caption{    \label{Fig:Al2O2} 
The two most stable isomers of Al$_2$O$_2$. The linear AlOAlO isomer is estimated to be 1,500 K higher in energy than the most stable (cyclic) isomer of Al$_2$O$_2$, but the rotational spectrum of the symmetric cyclic isomer is not observable.
 }
\end{center}
\end{figure}

\end{document}